\documentclass[journal]{IEEEtran}
\usepackage{cite}
\usepackage{amsmath,amssymb,amsfonts}
\usepackage{algorithmic}
\usepackage{graphicx,color}
\usepackage{textcomp}
\usepackage{xcolor}
\DeclareMathOperator{\Tr}{Tr}

\DeclareMathOperator*{\argmin}{arg\,min}
\usepackage[caption=false,font=footnotesize]{subfig}
\usepackage[short]{optidef}
\usepackage{bm}
\usepackage{comment}
\usepackage{hyperref}
\hypersetup{hidelinks=true}
\def\BibTeX{{\rm B\kern-.05em{\sc i\kern-.025em b}\kern-.08em
    T\kern-.1667em\lower.7ex\hbox{E}\kern-.125emX}}
\AtBeginDocument{\definecolor{tmlcncolor}{cmyk}{0.93,0.59,0.15,0.02}\definecolor{NavyBlue}{RGB}{0,86,125}}
\usepackage[linesnumbered,ruled,vlined]{algorithm2e}
\usepackage{setspace}
\def\BibTeX{{\rm B\kern-.05em{\sc i\kern-.025em b}\kern-.08em
    T\kern-.1667em\lower.7ex\hbox{E}\kern-.125emX}}
\SetKwInput{KwInput}{Input}                
\SetKwInput{KwOutput}{Output}              
\SetKw{KwInit}{Initialize}

\begin{document}

\title{Meta-Learning Based Optimization for Large Scale Wireless Systems}

\author{Rafael~Cerna~Loli,~\IEEEmembership{Member,~IEEE},
        and~Bruno~Clerckx,~\IEEEmembership{Fellow,~IEEE}
\thanks{R. Cerna Loli is supported by a grant provided by the Defence Science and Technology Laboratory (Dstl) Communications and Networks Research Programme.}
\thanks{R. Cerna Loli and B. Clerckx are with the Department of Electrical and Electronic Engineering, Imperial College London, London SW7 2AZ, U.K. (email: rafael.cerna-loli19@imperial.ac.uk; b.clerckx@imperial.ac.uk).}
}

\maketitle

\begin{abstract}
    Optimization algorithms for wireless systems play a fundamental role in improving their performance and efficiency. However, it is known that the complexity of conventional optimization algorithms in the literature often exponentially increases with the number of transmit antennas and communication users in the wireless system. Therefore, in the large scale regime, the astronomically large complexity of these optimization algorithms prohibits their use and prevents assessing large scale wireless systems performance under optimized conditions. To overcome this limitation, this work proposes instead the use of an unsupervised meta-learning based approach to directly perform non-convex optimization at significantly reduced complexity. To demonstrate the effectiveness of the proposed meta-learning based solution, the sum-rate (SR) maximization problem for the following three emerging 6G technologies is contemplated: hierarchical rate-splitting multiple access (H-RSMA), integrated sensing and communication (ISAC), and beyond-diagonal reconfigurable intelligent surfaces (BD-RIS). Through numerical results, it is demonstrated that the proposed meta-learning based optimization framework is able to successfully optimize the performance and also reveal unknown aspects of the operation in the large scale regime for the considered three 6G technologies.
\end{abstract}


\IEEEpeerreviewmaketitle

\section{Introduction}
\IEEEPARstart{T}{he} pathway towards 6G and beyond wireless technologies is now being actively researched and defined in both industry and academia. As a natural evolution of 5G New Radio (NR) and the current 5G Advanced technologies, 6G mobile communications are expected to bring a massive improvement in data throughput, spectral and energy efficiency, ultra reliability and minimal latency for critical services, global broadband for terrestrial and non-terrestrial communications, and other strict heterogeneous quality-of-service (QoS) constraints to guarantee extreme full-connectivity performance\cite{bruno_new, 6g_1, rs_fundamentals}. Specifically, 6G must achieve 50 times higher maximum data rates, 10 times reduced latency, and $10^4$ times higher reliability than 5G NR \cite{qos_6g} to realize, for instance, augmented/virtual reality (AR/VR) through further-enhanced mobile broadband (FeMBB), critical remote surgery and healthcare services through extremely ultra reliable and low-latency communications (eURLLC), and omnipresent internet-of-things through ultra massive machine type communications (umMTC). Complying with these rigorous constraints to enable a broad range of new sensing, localization, computation, powering, data-collection, and artificial intelligence (AI) functionalities is expected to ultimately lead to the creation of a new cyber-physical world \cite{ericsson_wp}.

To achieve this, 6G wireless networks must also implement other complex supporting technologies \cite{6g_isac, 6g_ris, 6g_2, 6g_3}. Among these, three in particular are expected to jointly play fundamental roles in 6G wireless networks: rate-splitting multiple access (RSMA), integrated sensing and communication (ISAC), and reconfigurable intelligent surfaces (RIS). Multiple access strategies are crucial in wireless communication systems in order to efficiently allocate the limited resources of the electromagnetic (EM) spectrum to simultaneously serve multiple users. In this context, RSMA has emerged recently as a powerful and flexible multi-user interference (MUI) management framework due to its capacity to partially decode interference and partially treat it as noise.  In particular, the superiority of RSMA over other multiple access techniques, such as space division multiple access (SDMA), which fully treats interference as noise, and non-orthogonal multiple access (NOMA), which fully decodes interference, lies in its ability to effectively adapt to varying degrees of network load, interference level and channel state information at the transmitter (CSIT) quality \cite{6g_1, eurasip, rs_fundamentals}. The efficient use of the limited EM spectrum is also the main motivation behind ISAC research. Specifically, the objective of ISAC systems is to employ unified hardware and signal processing units to jointly perform sensing and communication tasks, and, hence, improve the spectral efficiency and energy efficiency, and eliminate the inter-system interference compared to separately deploying radar and wireless communication systems \cite{6g_isac}. Finally, RIS-aided communications have emerged in recent years as an effort to shape the wireless channel to become more deterministic and, thus, facilitate achieving the service requirements at low operational costs. To achieve this, RISs would be deployed between the transmitter and the users to manipulate the impinging EM waves from the BS by employing passive scattering elements to reflect them towards the users after altering their amplitude and phase \cite{6g_ris}. Although the scattering elements in a RIS have been conventionally considered to be isolated with respect to each other, it has been recently demonstrated that interconnecting them brings superior performance at the expense of increased circuit complexity. This RIS structure has been denominated beyond-diagonal RIS (BD-RIS) \cite{hongyu_bd_ris}.

Naturally, optimization is a crucial step in order to improve the performance of a wireless system. However, optimizing the previously mentioned 6G supporting technologies is not a straightforward task due to their highly parameterized nature. Although various optimization algorithms have been devised in past works in the literature in order to improve different aspects of their performance, e.g. sum-rate (SR), energy efficiency, user-rate fairness, etc. \cite{eurasip, hongyu_bd_ris}, the main challenges of these conventional optimization algorithms are twofold:
\begin{enumerate}
    \item The optimization problems are typically non-convex and NP-hard. Thus, transformations of the objective function and its constraints are performed in order to convert the original non-convex and intractable problem into a convex and simpler, yet sub-optimal, approximation.
    \item Computationally costly operations, such as matrix inversions and eigendecompositions, are often employed in these conventional optimization techniques. Thus, their complexity increases rapidly with the deployment size, i.e. the number of antennas and users. This results in their application in large scale wireless systems being highly impractical.
\end{enumerate}

\subsection{Related works}
\label{literature_review}
As an effort to overcome the limitations of conventional optimization algorithms in wireless communications, Deep-Learning (DL) has received increased attention in recent years as a promising alternative due to its capacity to approximate complex functions with reduced computational complexity. Generally, DL-based optimization techniques can be classified into three main categories:

\textit{First,} the simplest approach is employing a \textit{black-box} generic neural network (NN) to approximate the variable of interest. For instance, in \cite{learn_opt_1} the authors employ a black-box deep NN (DNN) to approximate the classical weighted minimum mean squared error (WMMSE) precoder optimization algorithm for an SDMA system with perfect CSIT. Specifically, the DNN takes the vectorized channel matrix as the input and then it is trained using supervised learning, i.e. labeled training data for the DNN is generated by first running the conventional WMMSE precoder optimization algorithm, to maximize the system SR. This approach is also extended in \cite{learn_opt_2} to enable a robust WMMSE algorithm for massive MIMO SDMA systems with imperfect CSIT that maximizes the ergodic SR (ESR). In this case, the DNN is also trained in a supervised manner with a large dataset, albeit it outputs instead the Lagrange multipliers that are employed to compute the WMMSE precoders in closed form. The authors in \cite{learn_opt_3} focus on the hybrid precoder design for massive MIMO systems without explicit CSIT and employ a convolutional NN (CNN) trained in an unsupervised learning manner. The proposed CNN takes as input the quantized received signal strength indicator (RSSI) matrix of the system and it then simultaneously solves a regression problem to predict the fully digital precoders, and a classification problem to select the analog precoders from a pre-defined codebook.

\textit{Second,} exploiting the structure and known operations in an optimization algorithm can facilitate the design and reduce the complexity of a DL-based solution by appropriately selecting the size and number of the hidden layers. This approach is commonly named \textit{deep-unfolding} as each layer in the DNN aims to emulate one iteration of the optimization algorithm. A deep-unfolding solution of the WMMSE algorithm for a multi-user multiple-input-single-output (MU-MISO) system with perfect CSIT is proposed in \cite{learn_opt_4}. Essentially, the authors aim to bypass the costly matrix inversions, eigendecompositions, and bisection searches of the WMMSE algorithm by employing instead a projected-gradient approach which can be easily implemented using a DNN with a limited number of layers. The deep-unfolded DNN is then trained in an unsupervised manner to maximize the weighted SR over a large training data-set and their results show that the deep-unfolded solution can approximate the performance of the WMMSE algorithm as the number of layers increase. In \cite{learn_opt_5}, the same authors focus instead on MU-MIMO systems with perfect CSIT. By employing a gradient-descent strategy and Schulz iterations, a matrix-inversion-free variant of the WMMSE algorithm is created that can be readily used to implement a deep-unfolded solution. 

\textit{Third,} another type of DL-based precoder optimization solution is based on \textit{deep-reinforcement-learning} (DRL). The main premise of DRL-based optimization solutions is to treat the transmitter as the main agent whose policies and decisions on precoder selection have a direct effect on the environment (the communication system). After the agent has acted upon the environment, the latter returns a reward, in the form of the achievable system SR. Therefore, the agent employs a DNN in order to learn the optimum policies that maximize the reward. This process is iteratively refined through trial and error until the DNN converges to an optimum solution for the given channel conditions. In \cite{learn_opt_6}, a DRL-based solution is presented to perform power allocation and select the optimum precoders from a given codebook for each cell in a multi-cell network with imperfect CSIT. In \cite{learn_opt_7}, the use of DRL is explored in order to perform power allocation and select precoders to maximize the SR in a massive MU-MIMO system with imperfect CSIT. For RSMA systems, the authors in \cite{learn_opt_8} introduce a DRL-based framework to choose the precoders from a codebook and perform power allocation to maximize the SR in an aerial network with energy harvesting features.

\subsection{Motivation and main contributions}
DL-based optimization solutions face the following three limitations:
\begin{enumerate}
    \item The DL-based solutions often require a large training dataset in order to learn a generalized mapping rule to output precoder estimations for a large range of channel conditions and CSIT quality.
    \item If the DNNs are trained in a supervised learning manner, the performance of the estimated precoders is always upper-bounded by the underlying conventional optimization algorithm that was used to generate the training dataset. On the other hand, unsupervised learning is also used to train deep unfolding DNNs. In this case,  a single iteration of a conventional optimization algorithm is replaced by a single layer in the DNN. Hence, their performance is also upper-bounded by the underlying unfolded optimization algorithm, and by the number of layers (iterations) in its structure.
    \item Most DL-based works for precoder estimation consider the output of the DNN to be an estimated precoder matrix itself. Thus, this approach vastly increases the complexity of the DNN as its output space is large.
\end{enumerate}

To overcome the challenges faced by both conventional optimization algorithms and DL-based optimization solutions, meta-learning, a subset of machine learning (ML), has emerged very recently as a flexible and low-complexity alternative for optimization. At their core, meta-learning optimization solutions do not aim to learn or approximate an already existing optimization algorithm, but their objective is rather to \textit{learn how to learn} by repeatedly observing the training procedures of other conventional ML models, where each learns a separate task and without necessarily possessing a large training dataset. For instance, meta-learning techniques have been employed to optimize the precoders for SR maximization in \cite{learn_opt_9} for a single-user MIMO system with hybrid precoders. Specifically, a meta-learning model is trained to output a set of optimal neural network weights initialization to quickly adapt to unknown channel conditions. In \cite{lagd_1}, a meta-learning solution is proposed to maximize the system SR of a MU-MISO SDMA deployment with perfect CSIT. Numerical results demonstrate that the meta-learning precoder optimization solution outperforms the classical WMMSE algorithm by directly maximizing the non-convex SR expression instead of relying on convex relaxation techniques, which is the approach of the WMMSE algorithm. This strategy is also extended in \cite{lagd_2} for MU-MIMO SDMA systems with perfect CSIT. However, numerical results in \cite{lagd_1, lagd_2} are only provided for small scale deployments.

Given the lack of results to assess the optimized performance of large scale wireless systems, the objective of this paper is to directly address this issue by designing a generalized meta-learning based optimization framework for large scale wireless systems. The main contributions of this work are listed as follows:
\begin{itemize}
    \item First, this is the first work in the literature to demonstrate the effectiveness of meta-learning based optimization for highly-parameterized large scale wireless systems\footnote{In contrast with our previous work in \cite{rafael_meta}, this work proposes instead a generalized meta-learning based optimization framework for different large scale wireless systems.}. In particular, the work presented in this paper provides the first results in the literature for the SR maximization of hierarchical RSMA (H-RSMA) systems with imperfect CSIT and its extension to large scale ISAC systems, and the SR maximization of large scale fully connected BD-RIS systems with perfect CSIT. These three use cases are specifically chosen as they refer to the most advanced and generalized forms of multiple access schemes and intelligent surfaces architectures. \cite{rs_fundamentals, bruno_new, bruno_new_2}.
    \item Second, it is shown that by employing the proposed meta-learning based optimization framework, the intractable SR maximization problem for large scale wireless systems can be directly solved in an unsupervised learning manner and without relying on any convex relaxation techniques.
    \item Third, the meta-learning based optimization algorithm is shown to be able to optimize large scale wireless systems using the single CSIT realization as the unique datapoint in the training dataset, which stands in contrast with pure DL-based optimization methods that require massive training datasets to learn a specific optimization rule. Specifically, the proposed meta-learning based optimization framework exploits the overfitting effect of several compact DNNs, which are trained jointly, to directly learn to \textit{solve while training} \cite{lagd_1} for each CSIT realization. The number of DNNs is upper-bounded by the number of optimization variables in the optimization problem formulation.
    \item Fourth, the proposed meta-learning based optimization framework is designed so that its DNNs gradually learns to output the optimum gradient update with respect to fixed sub-optimal initialization points for each optimization variable. Differently from training a DNN to learn to output an actual precoder matrix itself, as evidenced from the related pure DL-based works in the literature mentioned in Section \ref{literature_review}, the approach used in this work effectively reduces the output space of the DNN as it is inherently tied to the fixed sub-optimal initialization point of the optimization variable of interest. Thus, the use of an appropriate initialization point to ensure optimization convergence is also necessary, similarly to what is required in conventional optimization algorithms.
    \item Fifth, the effectiveness of the meta-learning based optimization framework is demonstrated through numerical results\footnote{The meta-learning based optimization framework is implemented in Python 3.9.16 with PyTorch 2.0.1, using an NVIDIA RTX 6000 GPU.}. For H-RSMA with imperfect CSIT, it is shown that the meta-learned precoders vastly outperform other low-complexity precoders in the literature. Also, the gains of H-RSMA over SDMA when complying with QoS rate constraints in the large scale regime are revealed. Specifically, it is shown that when the CSIT quality is severely degraded and the user channels are highly correlated, the ESR gain of H-RSMA over SDMA in the high SNR regime is over 50\%. For large scale ISAC, it is revealed that the global common stream precoder of H-RSMA is the fundamental factor to achieve an ESR gain of over 300\% compared to SDMA when fully prioritizing radar sensing functions. For large-scale BD-RIS, the use of the devised meta-learning algorithm confirms that the SR of fully-connected BD-RIS is far superior than the SR attainable by conventional RIS in the large scale regime.
\end{itemize}

\textit{Notation:} Scalars, vectors and matrices are denoted by standard, bold lower and upper case letters, respectively. $\mathbf{I}$ denotes the identity matrix. The Hermitian transpose operator and matrix trace operator are represented by $(.)^H$, and $\text{Tr}(.)$, respectively. The expectation of a random variable is given by $\mathbb{E}\{.\}$. $||.||_2$ is the $l_2$-norm operator, and $\max(.,.)$ is the operator that returns the maximum between its input parameters.  Also, $\mathtt{\sim}$ denotes ``distributed as" and $\mathcal{CN}(0,\sigma^2)$ denotes
the Circularly Symmetric Complex Gaussian (CSCG) distribution with zero mean and variance $\sigma^2$. 

\section{Fundamentals of meta-learning}
\label{meta_fundamentals}
DL-based solutions can be successful in learning inference rules that rival or even outperform classical model-based methods in wireless communications. However, they often require large training datasets to learn a specific mapping, even after using deep-unfolding. Also, in the case of DRL models, they usually need a considerable number of time-slots to converge due to the numerous agents in the wireless communication system. This data dependency behaviour of DNNs is expected as they are trained to learn feature representation. In other words, they learn complex underlying structures from the data points in the training dataset in order to implement an inference rule.

The human learning process is vastly different from that of DL. Specifically, humans are often able to learn and generalize concepts quickly after a single (or a few instances of) exposure. Even more, human intelligence is trained for a lifetime, and learning skills evolve with experience, allowing humans to train a unique \textit{learning to learn} ability. This is the main objective of meta-learning research \cite{meta_1, meta_2,meta_3}. 
\begin{figure}[t!]
		\centering
        \includegraphics[width=\columnwidth]{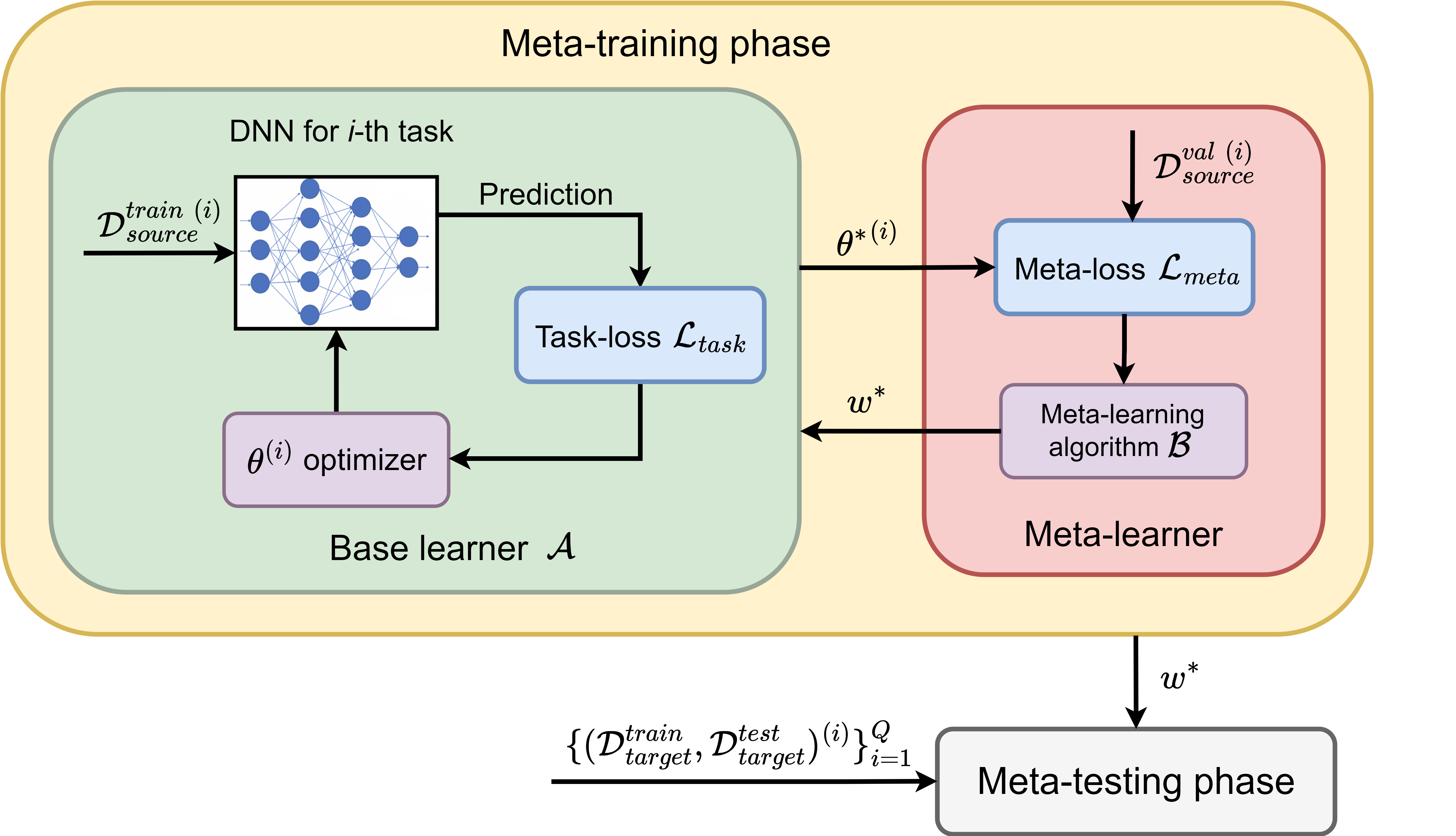}
		\caption{General structure of the meta-learning framework.}
		\label{fig:meta_learning_model}
\end{figure}
The general structure of the meta-learning framework is then depicted in Fig. \ref{fig:meta_learning_model}, where it is observed that it involves a meta-training phase, in which the meta-learning algorithm $\mathcal{B}$ learns how to learn by observing the learning process of a base learner $\mathcal{A}$ to solve a set of $M$ source tasks, and a meta-testing phase, in which the performance of the meta-learning algorithm $\mathcal{B}$ is evaluated. This is done by assessing the effectiveness of its output, the meta-learned knowledge $w^*$, in enabling the base learner $\mathcal{A}$ to generalize across $Q$ previously unseen target tasks. The meta-knowledge $w^*$ may consist of different parameters that affect a learning algorithm, e.g. an optimized estimate of the initial DNN parameters of the base learner $\mathcal{A}$ \cite{meta_3}, an optimization strategy \cite{meta_4}, a meta-learned loss function to jointly update the parameters of the DNNs in the base learner, or even an entire learning model, such as an optimized DNN structure (number of layers, neurons, activation functions, etc).

Specifically, in the meta-training phase, $M$ source tasks are drawn from a task distribution $p(\mathcal{T})$, and the total dataset associated with them is denoted $\mathcal{D}_{source}^{total} = \{(\mathcal{D}_{source}^{train},\mathcal{D}_{source}^{val})^{(i)}\}_{i=1}^M$, which contains training and validation data points. The meta-training phase consists of an inner layer and an outer layer that iteratively exchange updated parameters between them. In the inner layer, the base learner $\mathcal{A}$, which could be a DNN as depicted in Fig. \ref{fig:meta_learning_model}, is trained for a number of inner iterations to minimize the task-loss function $\mathcal{L}_{task}$ for each of the $M$ tasks, using the meta-knowledge $w$ obtained previously from the meta-learner and the corresponding training data points, in order to compute the optimum parameters $\{{\bm{\theta}^*}^{(i)}\}_{i=1}^M$. These optimized parameters are then used in the outer layer to train the meta-learning algorithm $\mathcal{B}$ to minimize the meta-loss function $\mathcal{L}_{meta}$, which describes the cumulative generalization error over the validation training data points in the meta-training dataset for each of the $M$ source tasks.  The meta-learning algorithm $\mathcal{B}$ then outputs the refined meta-knowledge $w^*$ that minimizes $\mathcal{L}_{meta}$ and returns it to the base learner in the inner layer to repeat the process in the next iteration. Thus, the meta-training phase can be described as a bilevel optimization problem \cite{meta_2}, i.e. an optimization problem that has another optimization as constraint, given by
\begin{align}
    w^* &= \argmin_{w}\sum_{i=1}^M\mathcal{L}_{meta}(\mathcal{D}^{val\;(i)}_{source};{\bm{\theta}^*}^{(i)}(w)) \label{meta_outer}\\
    \text{s.t.}\;\;\;\;&\;\;{\bm{\theta}^*}^{(i)}(w)=\argmin_{\bm{\theta}}\;\mathcal{L}_{task}(\mathcal{D}^{train\;(i)}_{source};{\bm{\theta}},w).
\end{align}

Next, consider a set of unseen $Q$ target tasks to evaluate the performance of the meta-learning algorithm $\mathcal{B}$ in the meta-testing phase. Given the testing dataset $\mathcal{D}_{target}^{total} = \{(\mathcal{D}_{target}^{train},\mathcal{D}_{target}^{test})^{(i)}\}_{i=1}^Q$, first the base learner $\mathcal{A}$ is trained using the optimized meta-learned knowledge $w^*$ on each of the $Q$ target tasks as follows
\begin{equation}
    {\bm{\theta}^*}^{(i)}=\argmin_{\bm{\theta}}\;\mathcal{L}(\mathcal{D}^{train\;(i)}_{target};{\bm{\theta}},w^*).
\end{equation}
The accuracy of the meta-learning algorithm is then evaluated by the performance of using the learned ${\bm{\theta}^*}^{(i)}$ on the testing data points $\mathcal{D}^{test\;(i)}_{target}$ for each of the $Q$ target tasks. Finally, to summarize from the description of the meta-learning framework, its core three components are highlighted \cite{meta_2}:
\begin{itemize}
    \item \textit{Meta-objective}: The \textit{``Why?"} of the meta-learning algorithm. This covers the choice of meta-loss function $\mathcal{L}_{meta}$ to be minimized, and the task distribution $p(\mathcal{T})$ from which the $M$ source tasks are selected.
    \item \textit{Meta-optimizer}: The \textit{``How?"} of the the meta-learning algorithm. That is, the choice of optimizer to use in the outer layer during meta-training to solve (\ref{meta_outer}).
    \item \textit{Meta-representation}: The \textit{``What?"} of the meta-learning algorithm. In other words, what meta-knowledge $w$ needs to be learned from observing the learning processes of the base learner for the $M$ source tasks.
\end{itemize}

\subsection{Example}
A simple example to illustrate the application of meta-learning in the context of wireless communications is given next. Consider a multi-antenna transmitter with imperfect CSIT that communicates with multiple users. A meta-learner can be trained to learn how to optimize the precoders, similarly to the work in \cite{learn_opt_9}, by observing the training procedure of a base learner, which employs a DNN to solve a set of $M$ SR maximization source tasks, each with varying levels of CSIT quality. Thus, the training and validation data for the $i$-th source task $(\mathcal{D}_{source}^{train},\mathcal{D}_{source}^{val})^{(i)}$ may be composed of different CSIT realizations associated with the given CSIT quality. If the DNN in the base learner is trained in an unsupervised learning manner, the task-loss function can be given by
\begin{equation}
    \mathcal{L}_{task}(\mathcal{D}^{train\;(i)}_{source};{\bm{\theta}},w)=\frac{1}{n_t^{(i)}}\sum_{n=1}^{n_t^{(i)}}\text{SR}_n({\bm{\theta}},w),
\end{equation}
where $\bm{\theta}$ denotes the network weights of the DNN, and $w$ can be an specific initialization of the network weights obtained from the meta-learner in the last iteration of the outer layer. Also, $n_t^{(i)}$ denotes the number of training data points in $\mathcal{D}^{train\;(i)}_{source}$, and $\text{SR}_n({\bm{\theta}},w)$ is the SR achieved for the $n$-th training data point using $\bm{\theta}$, and $w$. After all different precoder optimization tasks are solved in the inner layer by the base learner, the meta-learner retrieves the updated network weights for all tasks and calculates the meta-loss for each task. The meta-loss for source task-$i$ can be given by
\begin{equation}
    \mathcal{L}_{meta}(\mathcal{D}^{val\;(i)}_{source};{\bm{\theta}^*}^{(i)}(w))=\frac{1}{n_v^{(i)}}\sum_{n=1}^{n_v^{(i)}}\text{SR}_n({\bm{\theta}^*}^{(i)}(w)),
\end{equation}
where, analogously, $n_v^{(i)}$ denotes the number of validation data points in $\mathcal{D}^{val\;(i)}_{source}$. Across outer iterations, the meta-knowledge $w$ is then updated, for instance, using a gradient descent method until convergence is reached to obtain the refined meta-knowledge $w^*$ \cite{meta_2}. In turn, $w^*$ can then be used as an optimized initialization of the DNN weights in the base learner when training it to optimize the precoders for new channel conditions in which large training datasets are not necessarily available.

\section{Meta-learning based optimization framework}
In this section, the design of the proposed meta-learning based non-convex optimization framework is presented. First, a generalized system model for large scale wireless systems is described. Then, the design of the meta-learning optimization algorithm is discussed.

\subsection{A generalized system model}
It is important to emphasize that the proposed meta-learning based optimization framework is not designed only for a specific application. Therefore, consider a generic wireless communication system, in which a transmitter equipped with an arbitrarily large number of transmit antennas, i.e. $N_t \gg 1$, serves numerous communication users $K \gg 1$ using $D$ different modulated data streams given by 
\begin{equation}
    \mathbf{s} = [\mathbf{s}_c^T, \mathbf{s}_p^T]^T  = [\underbrace{s_{c,1},\dots,s_{c,M}}_{M \text{ streams}},\underbrace{s_{p,1},\dots,s_{p,K}}_{K \text{ streams}}]\in \mathbb{C}^{D \times 1}, 
\end{equation}
where $\mathbf{s}_c  \in \mathbb{C}^{M \times 1}$ contains the $M$ streams that are decoded by more than one user; and $\mathbf{s}_p  \in \mathbb{C}^{K \times 1}$, the $C$ streams decoded only by a specific user. Before transmission, the data streams are precoded to generate the transmit signal as follows
\begin{equation}
    \mathbf{x} = \mathbf{P}\mathbf{s} = [\underbrace{\mathbf{p}_{c,1},\dots,\mathbf{p}_{c,M}}_{M \text{ precoders}},\underbrace{\mathbf{p}_{p,1},\dots,\mathbf{p}_{p,K}}_{K \text{ precoders}}]\mathbf{s} \in \mathbb{C}^{N_t \times 1},
\end{equation}
where $\mathbf{P} \in \mathbb{C}^{N_t \times D}$ is the precoder matrix. Therefore, a generic rate-related optimization problem can be expressed as follows
\begin{maxi!} |s|[2]
{\mathbf{X}_1,\dots,\mathbf{X}_N}{f(R_{c,1},\dots,R_{c,M},R_{p,1},\dots,R_{p,K})}{\label{general_opt_prob}}{} 
\addConstraint{\mathbb{E}\{\Tr(\mathbf{x}\mathbf{x}^H)\}}{\leq P_t,\label{eq:C1_general}}
\addConstraint{\quad\quad\quad\quad\vdots}{}
\end{maxi!}
where the term $R_i$ denotes the achievable rate related to the data stream with index $i$, and $\mathbf{X}_1,\dots,\mathbf{X}_N$ are the $N$ independent variables that are jointly optimized to maximize the objective function $f(.)$. The latter may consider rate-related metrics, e.g. SR, max-min fairness (MMF), etc., as they are generally considered representative of the performance of wireless communication systems. Naturally, the precoder matrix $\mathbf{P}$ may be one of the optimization variables $N$ optimization variables as the achievable rates of each user directly depend on it. Alternatively, it is also possible to consider precoders with fixed direction and optimize a precoder power allocation vector instead with lower complexity. 

\subsection{Meta-learning based optimization framework design}
Given the generic optimization problem in (\ref{general_opt_prob}), the 
design of a meta-learning based optimization framework can be approached by first defining, as explained in Section \ref{meta_fundamentals},  its three core meta-knowledge components as follows:
\begin{itemize}
    \item \textit{Meta-objective:} The meta-learning based optimization framework is not designed to be used to solve a set of different tasks, but rather it is particularly tuned to focus on solving the optimization problem in (\ref{general_opt_prob}). Thus, it can be classified as a single-task meta-learning algorithm. The meta-loss function $\mathcal{L}_{meta}$ can be defined by modifying the original objective function $f(.)$ to include additional optimization constraints using appropriately selected regularization parameters to enforce them.
    \item \textit{Meta-optimizer:} As demonstrated in \cite{mlam,meta_3}, the use of a gradient descent meta-optimizer has found significant success in solving non-convex optimization problems using a meta-learning approach. Thus, the proposed solution also contemplates its implementation.
    \item \textit{Meta-representation:} The output of the meta-learning based optimization framework is directly the set of optimum parameters $\{\mathbf{X}_1^*,\dots,\mathbf{X}_N^*\}$ that minimizes best the meta-loss function $\mathcal{L}_{meta}$.
\end{itemize}
\begin{figure}[t!]
		\centering		        
        \includegraphics[width=\columnwidth]{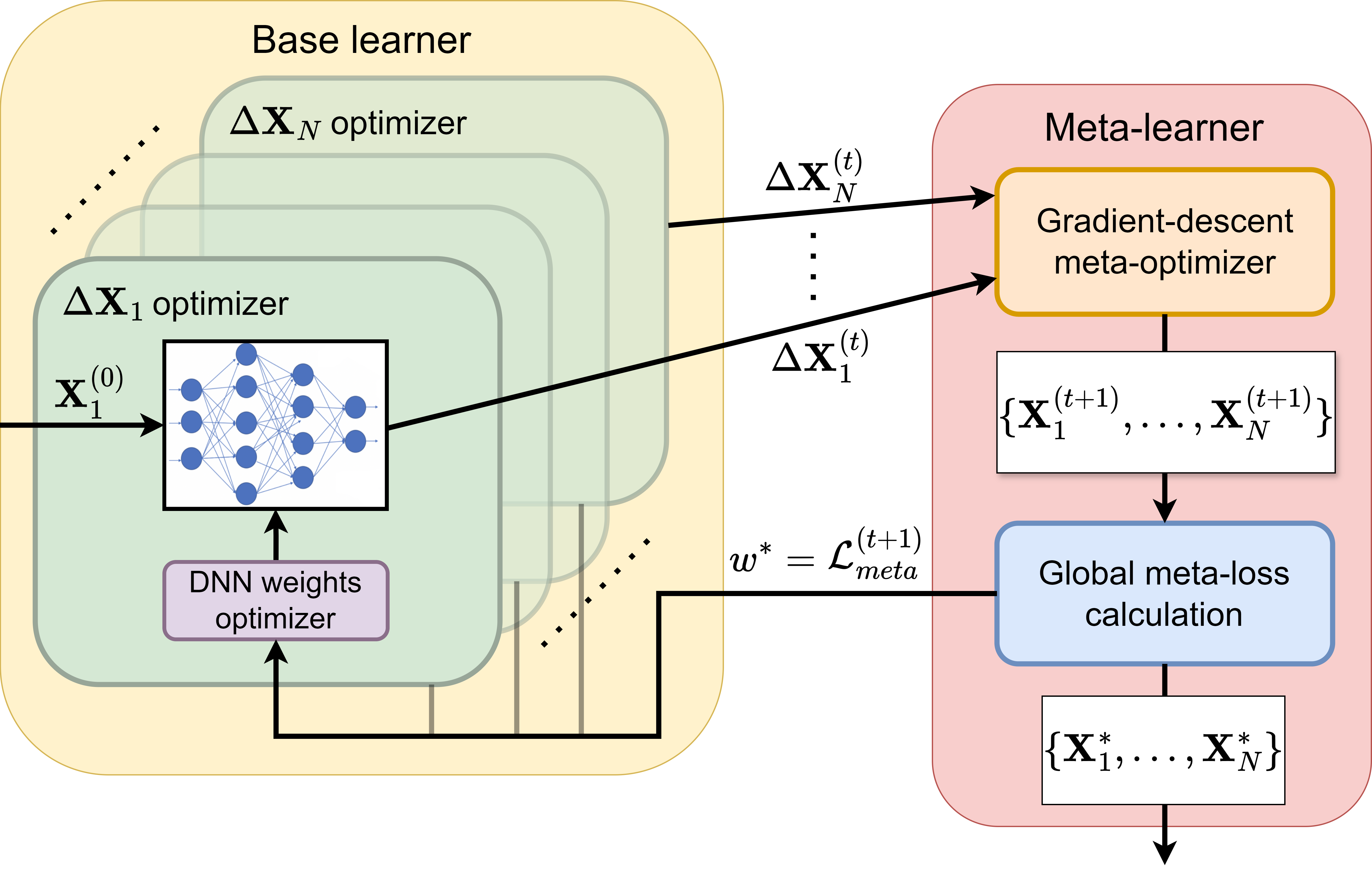}
		\caption{Proposed meta-learning based optimization algorithm structure.}
		\label{fig:algo_structure}
\end{figure}

Given these considerations, a high-level structure of the devised meta-learning optimization algorithm is then depicted in Fig. \ref{fig:algo_structure}, and its detailed operation is described next. During the $t$-th iteration of the meta-learning based optimization process, a base learner employs a set of $N$ DNNs to predict a set of the next gradient updates $\{\mathbf{X}_1^{(t)},\dots,\mathbf{X}_N^{(t)}\}$ (each DNN outputs the gradient update for a specific optimization variable). The base learner forwards the predicted set of gradient updates to the meta-learner, which employs the gradient descent meta-optimizer to separately update each optimization variable as follows
\begin{equation}
    \mathbf{X}_i^{(t+1)} = \mathbf{X}_i^{(0)} + \Delta\mathbf{X}_i^{(t)}\;,\; \forall i \in \{1,\dots,n\}, 
\end{equation}
where  $\mathbf{X}_i^{(0)}$ is a fixed sub-optimal estimation of the $i$-th optimization variable (if $\mathbf{X}_i^{(0)} \in \mathbb{C}$, it can be formatted as $[\Re\{\mathbf{X}_i^{(0)}\}, \Im\{\mathbf{X}_i^{(0)}\}]$) After the $N$ optimization variables have been updated, the meta-loss is calculated, and if a lower meta-loss is achieved compared to the previous iteration, the meta-learner stores it as the optimum values. Finally, the updated meta-loss is returned to the base learner so that it can employ it to update the weights of each DNN in a separate manner. This process is repeated for a total of $T$ iterations after which the meta-learner outputs its estimation of the optimum values $\{\mathbf{X}_1^*,\dots,\mathbf{X}_N^*\}$. The following observations are then made: 
\begin{itemize}
    \item Although the works in \cite{mlam, meta_3} considered the use of recurrent neural networks (RNNs) due to their ability to memorize and forget sequential information, it was demonstrated in \cite{lagd_1, lagd_2} that using a single DNN can provide better performance at a lower complexity through using an Adam optimizer, which stores a rolling average of previous gradients used to optimize the DNN weights. Thus, the proposed meta-learning solution also uses  this structure.
    \item Unlike the framework in \cite{mlam}, which trains the RNNs in the base learner using several inner iterations (epochs), the proposed solution considers instead that the base learner uses only a single training epoch before outputting the gradient update. As stated in \cite{meta_2}, meta-optimization through gradient descent over a long number of inner iterations leads to compute and memory issues. Thus, this is avoided in the proposed solution.
    \item Differently from \cite{lagd_1, lagd_2}, the inputs to the DNNs in every iteration are fixed to the initial sub-optimal estimations of the $N$ optimization variables, denoted by $\{\mathbf{X}_1^{(0)}, \dots, \mathbf{X}_N^{(0)}\}$. This is done in order to simplify the learning task of each DNN compared to requiring it to learn to output the next optimum gradient for each updated value of the optimization variable. Alternatively, this can be understood as effectively employing a training dataset of only one datapoint to  achieve a reduction in training time and DNN complexity, while also exploiting the overfitting effect of each DNN to the initial input to achieve \textit{solving while training}.
    \item The $N$ DNNs can be trained in parallel. Therefore, the complexity of the proposed meta-learning based optimization framework is $\mathcal{O}(TAB)$ where $A$ and $B$ denote the number of rows and columns of the optimization variable with the largest dimensions.
\end{itemize}

\subsection{Single-variable meta-learning based algorithm for precoder optimization}

\begin{algorithm}[t!]
\DontPrintSemicolon
  \KwInput{$N_t, K, \hat{\mathbf{H}}, T, \beta.$}
   
  \text{Initialize} $\mathbf{P}_0$, $\bm{\theta}_0$

   $\mathcal{L}^{*}_{meta} = \mathcal{L}_{meta}(\mathbf{P}_0)$
  
  \For{$t \leftarrow 0,1,\dots,T-1$}
  {$\Delta\mathbf{P}_t = \text{G}_{\bm{\theta}_t}(\nabla_{\mathbf{P}_0}\mathcal{L}_{meta}(\mathbf{P}_0))$\;
  $\mathbf{P}_{t+1} = \mathbf{P}_0 + \Delta\mathbf{P}_t$ \;
  $\mathbf{P}_{t+1} = \Omega(\mathbf{P}_{t+1})$\;
  \uIf{$\mathcal{L}_{meta}(\mathbf{P}_{t+1})<\mathcal{L}^{*}_{meta}$}{
    $\mathcal{L}^{*}_{meta} = \mathcal{L}_{meta}(\mathbf{P}_{t+1})$ \;
    $\mathbf{P}^{*} = \mathbf{P}_{t+1}$ \;
  }
  $\Delta\bm{\theta}_t = \beta \cdot \text{Adam}(\nabla_{\bm{\theta}_{t}}\mathcal{L}^{meta}(\mathbf{P}_{t+1}))$\;
  $\bm{\theta}_{t+1} = \bm{\theta}_{t} +\Delta\bm{\theta}_t$ \;
  }
  \KwOutput{$\mathcal{L}^{*}_{meta},\mathbf{P}^{*}$}
\caption{Single-Variable Meta-Learning Based Algorithm for Precoder Optimization}
\label{meta_alg_1}
\end{algorithm}

Consider first the simplest and most common application: the single variable meta-learning based framework for precoder optimization. This is summarized in Algorithm \ref{meta_alg_1} and a detailed description is given next.

At the start of the meta-learning based precoder optimization algorithm, an initial sub-optimal precoder matrix $\mathbf{P}_0$ is computed as a function of the available CSIT $\hat{\mathbf{H}}$. Also, the tunable parameters $\bm{\theta}$ of the DNN $\text{G}_{\bm{\theta}}(.)$ in the base learner are initialized as $\bm{\theta}_0$. At the $t$-th iteration, the DNN $\text{G}_{\bm{\theta}_t}(.)$ takes as the fixed 
 input the gradient with respect to $\mathbf{P}_0$ of the meta-loss function, denoted by $\nabla_{\mathbf{P}_0}\mathcal{L}_{meta}(\mathbf{P}_0)$. The output of $\text{G}_{\bm{\theta}_t}(.)$ is next used to compute the incremental precoder update term as follows  
\begin{equation}
    \mathbf{P}_{t+1} = \mathbf{P}_0 + \text{G}_{\bm{\theta}_t}(\nabla_{\mathbf{P}_0}\mathcal{L}_{meta}(\mathbf{P}_0)).
\end{equation}
The operator $\Omega(.)$ is then employed to project the updated precoder matrix $\mathbf{P}_{i+1}$ and enforce the total transmit power constraint in ($\ref{eq:C1_general})$ as follows
\begin{equation}
\Omega(\mathbf{P}) = 
\begin{cases}
        \mathbf{P}, & \text{if } \Tr(\mathbf{P}\mathbf{P}^H)\leq P_t\\
        \sqrt{\frac{P_t}{\Tr({\mathbf{P}\mathbf{P}^H})}}\mathbf{P} & \text{otherwise.}
    \end{cases},
\end{equation}
where the transmit power constraint is considered as $\Tr(\mathbf{P}\mathbf{P}^H)\leq P_t$ instead of $\mathbb{E}\{\mathbf{x}\mathbf{x}^H)\}\leq P_t$, assuming that $\mathbb{E}\{\mathbf{s}\mathbf{s}^H\} = \mathbf{I}$. If the new meta-loss value $\mathcal{L}_{meta}(\mathbf{P}_{t+1})$ is lower than the previous optimum meta-loss $\mathcal{L}^{*}_{meta}$, then $\mathcal{L}^{*}_{meta}$ and the optimum precoder matrix $\mathbf{P}^*$ are also updated and buffered. Finally, the tunable parameters $\bm{\theta}$ are updated using the Adam \cite{adam} optimizer with respect to the new meta-loss value $\mathcal{L}_{meta}(\mathbf{P}_{i+1})$ as follows \cite{lagd_2}
\begin{equation}
    \bm{\theta}_{i+1} = \bm{\theta}_{i} + \beta \cdot \text{Adam}(\nabla_{\bm{\theta}_{i}}\mathcal{L}(\mathbf{P}_{i+1})),
\end{equation}
where $\beta$ is the learning rate of the Adam optimizer. After the $I$ iterations, the meta-learning algorithm outputs $\mathbf{P}^*$ and $\mathcal{L}^*_{meta}$.

\section{Use case 1: Hierarchical rate-splitting multiple access}
\label{hrsma_section}
\subsection{System model}
\begin{figure}[t!]
		\centering		        
        \includegraphics[width=0.95\columnwidth]{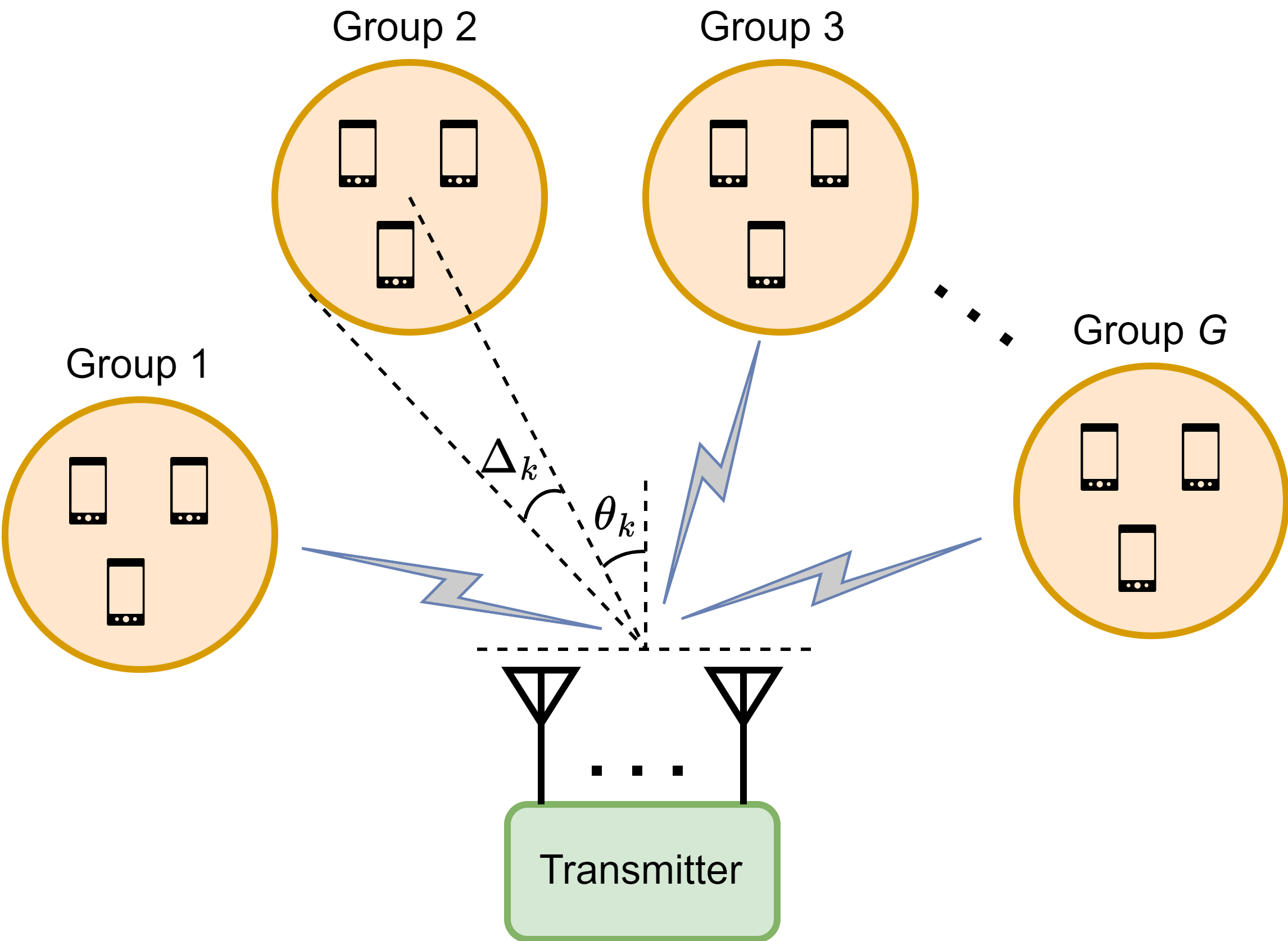}
		\caption{H-RSMA system model.}
		\label{fig:hrs_model}
\end{figure}
Consider a transmitter equipped with $N_t$ transmit antennas that serves $K$ single-antenna communication users, indexed by the set $\mathcal{K}=\{1,\dots,K\}$, in the downlink direction as depicted in Fig. \ref{fig:hrs_model}. Assuming a geometrical one-ring scattering model \cite{one_ring}, the correlation between the channel coefficients of transmit antennas $1\leq i,j \leq N_t$ can be expressed as follows:
\begin{equation}
    [\mathbf{R}_k]_{i,j} = \frac{1}{2\Delta_k}\int_{\theta_k-\Delta_k}^{\theta_k+\Delta_k} e^{-j\frac{2\pi}{\lambda}\Psi(\alpha)(\mathbf{r}_i-\mathbf{r}_j)}d\alpha,
    \label{corr_matrix}
\end{equation}
where $\theta_k$ is the azimuth angle towards user-$k$, $\Delta_k$ denotes the angular spread of departure to user-$k$, $\Psi(\alpha) = [\cos{\alpha}, \sin{\alpha}]$ is the wave vector for a planar wave arriving at the $\alpha$ angle, and $\mathbf{r}_i = [x_i, y_i]^T$ is the position vector of antenna $i$. The CSI model for user-$k$ is then given by 
\begin{equation}
\begin{split}    
    \mathbf{{h}}_k &=  \sqrt{1-\sigma_{e,k}^2}\mathbf{\hat{h}}_k + \sigma_{e,k}\Tilde{\mathbf{h}}_k \\ &= \mathbf{R}_k^{\frac{1}{2}}\mathbf{g}_k = \mathbf{R}_k^{\frac{1}{2}}\Big(\sqrt{1-\sigma_{e,k}^2}\mathbf{\hat{g}}_k+\sigma_{e,k}\mathbf{z}_k\Big),
\end{split}
    \label{partial_csit_model}
\end{equation}
where $\mathbf{{h}}_k \in \mathbb{C}^{N_t \times 1}$ is the real CSI for user-$k$, $\mathbf{\hat{h}}_k \in \mathbb{C}^{N_t \times 1}$ is the CSIT, and $\Tilde{\mathbf{h}}_k \in \mathbb{C}^{N_t \times 1}$ is the CSIT error. Also, $\mathbf{g}_k \in \mathbb{C}^{N_t \times 1}$, $\mathbf{\hat{g}}_k \in \mathbb{C}^{N_t \times 1}$ and $\mathbf{z}_k \in \mathbb{C}^{N_t \times 1}$ have i.i.d entries drawn from the distribution $\mathcal{CN}(0,1)$, and $\sigma_{e,k}^2 \in [0, 1]$ denotes the CSIT error variance for user-$k$. The system CSI, CSIT, and CSIT error are then denoted respectively by $\mathbf{H}=[\mathbf{h}_1,\dots,\mathbf{h}_K]$ $\hat{\mathbf{H}}=[\hat{\mathbf{h}}_1,\dots,\hat{\mathbf{h}}_K]$, and $\Tilde{\mathbf{H}}=[\Tilde{\mathbf{h}}_1,\dots,\Tilde{\mathbf{h}}_K]$.

It is further considered that the $K$ users are clustered into $G$ groups as also shown in Fig. \ref{fig:hrs_model}, indexed by the set $\mathcal{G}=\{1,\dots, G\}$, according to the similarity degree of their spatial correlation matrices $\mathbf{R}_k\in \mathbb{C}^{N_t \times N_t},\forall k \in \mathcal{K}$. Then, for a given group-$g$, we denote the number of users in it by $K_g$, and consider them to be internally indexed by the set $\mathcal{K}_g=\{1,\dots, K_g\}$. In H-RSMA, the message of user-$k$, $W_k$, is split at the transmitter into a global common part $W_{c,k}$, a group common part $W_{c,g,k}$, and a private part $W_{p,k}$, $\forall k \in \mathcal{K}$. Then, all the global common parts of the $K$ users $\{W_{c,1}, \dots, W_{c,K}\}$ are jointly encoded and modulated into a single global common stream $s_c$. The group common parts of the $K_g$ users in group-$g$ are also jointly encoded and modulated into the group common stream $s_{c,g}$, $\forall g \in \mathcal{G}$. Finally, the private parts $\{W_{p,1}, \dots, W_{p,K}\}$ are encoded and modulated independently into $K$ private streams $\{s_1, \dots, s_K\}$. All streams are next linearly precoded using the precoder matrix $\mathbf{P} = [\mathbf{p}_c,\mathbf{p}_{c,1},\dots,\mathbf{p}_{c,G},\mathbf{p}_1,\dots,\mathbf{p}_K] \in \mathbb{C}^{N_t \times (K+G+1)}$, where $\mathbf{p}_c$ is the global common precoder, $\mathbf{p}_{c,g}$ is the group common precoder for group-$g$, and $\mathbf{p}_k$ is the private precoder for user-$k$. The transmitted signal $\mathbf{x} \in \mathbb{C}^{N_t \times 1}$ is given by
\begin{equation}
    \mathbf{x} = \mathbf{P}\mathbf{s} = \mathbf{p}_cs_c + \sum_{g=1}^{G}\mathbf{p}_{c,g}s_{c,g} + \sum_{k=1}^{K}\mathbf{p}_ks_k,    
    \label{hrsma_tx_signal}
\end{equation}
where $\mathbf{s} = [s_c,s_{c,1},\dots,s_{c,G},s_1,\dots,s_K]^T \in \mathbb{C}^{(K+G+1)\times 1}$. It is assumed that $\mathbb{E}\{\mathbf{s}\mathbf{s}^H\}=\mathbf{I}_{(K+G+1)}$ and, hence, the total transmit power constraint is expressed as $\Tr(\mathbf{P}\mathbf{P}^H)\leq P_t$. The received signal at the output of the antenna of user-$k$, which belongs to group-$g$, is then given by
\begin{equation}
    \begin{split}
        y_k &= \mathbf{h}_k^H\mathbf{p}_cs_c+ \mathbf{h}_k^H\mathbf{p}_{c,g}s_{c,g}+\mathbf{h}_k^H\mathbf{p}_k s_k+\\
        &\;\;\;\underbrace{\sum_{n\neq g}^G\mathbf{h}_k^H\mathbf{p}_{c,n} s_{c,n}}_{\text{inter-group interference}}+\underbrace{\sum_{j\neq k}^K\mathbf{h}_k^H\mathbf{p}_j s_j}_{\text{multi-user interference}}+n_k,
    \end{split}
\end{equation}
where $\mathbf{h}_k \in \mathbb{C}^{N_t \times 1}$ is the downlink channel between the transmitter and user-$k$, and $n_k\;\mathtt{\sim}\; \mathcal{CN}(0,\sigma_{n,k}^2)$ is the AWGN at the antenna of user-$k$.

At user-$k$, the decoding operations are performed as follows assuming perfect CSIR. First, the global common stream $s_c$ is decoded by treating all other streams as noise. User-$k$ then employs SIC to subtract the interference from the global common stream from the received signal $y_k$, and then attempts to decode its group common message $s_{c,g}$ by treating the rest of the group common and private streams as noise. Finally, user-$k$ employs SIC again to remove the interference from the group common message from the remaining signal, and decodes its own private message $s_k$. The SINRs at user-$k$ of decoding $s_c, s_{c,g}$ and $s_k$, respectively, are given by
\begin{equation}
    \begin{split}
        \gamma_{c,k} &= \frac{|\mathbf{h}_k^H\mathbf{p}_c|^2}{\sum_{n\in\mathcal{G}}^G|\mathbf{h}_k^H\mathbf{p}_{c,n}|^2+\sum_{j\in\mathcal{K}}^K|\mathbf{h}_k^H\mathbf{p}_j|^2+\sigma_{n,k}^2},\\
        \gamma_{c,g,k} &= \frac{|\mathbf{h}_k^H\mathbf{p}_{c,g}|^2}{\sum_{n\neq g}^G|\mathbf{h}_k^H\mathbf{p}_{c,n}|^2+\sum_{j\in\mathcal{K}}^K|\mathbf{h}_k^H\mathbf{p}_j|^2+\sigma_{n,k}^2},\\
        \gamma_{p,k} &= \frac{|\mathbf{h}_k^H\mathbf{p}_k|^2}{\sum_{n\neq g}^G|\mathbf{h}_k^H\mathbf{p}_{c,n}|^2+\sum_{j\neq k}^K|\mathbf{h}_k^H\mathbf{p}_j|^2+\sigma_{n,k}^2}.
    \end{split}
    \label{sinr_expressions}
\end{equation}
Considering Gaussian signalling, the achievable rate of the global common stream at user-$k$ is $R_{c,k} = \log_2(1+\gamma_{c,k})$, the achievable rate of the group common stream of group-$g$ at user-$k$ is given by $R_{c,g,k}=\log_2(1+\gamma_{c,g,k})$, and the achievable rate of the private stream of user-$k$ is $R_{k} = \log_2(1+\gamma_{p ,k})$. As the global common stream and group common streams must be decoded by more than one user, they must be transmitted, respectively, at rates lower than $R_c \leq \min\{R_{c,1},\dots,R_{c,K}\}$ and $R_{c,g} \leq \min\{R_{c,g,1},\dots,R_{c,g,K_g}\}, \forall g \in \mathcal{G}$.

\subsection{Problem formulation}
Due to imperfect CSIT, it is not possible to determine the exact instantaneous rates of the common and private streams. A naive strategy would be then to optimize the precoder matrix $\mathbf{P}$ using the imperfect CSIT as the real CSI. However, a more robust approach is to aim to maximize the ergodic SR (ESR) instead, as proposed in \cite{hamdi}. The ESR expression is given by
\begin{equation}
    \text{ESR}(\mathbf{P}) = \mathbb{E}_{\{\text{H},\hat{\text{H}\}}}\Bigg\{R_c+\sum_{g=1}^G{R}_{c,g}+\sum_{k=1}^K{R}_{p,k}\Bigg\}.
    \label{esr_1lrs}
\end{equation}
The ESR expression cannot be directly optimized with ease as it entails maximizing (\ref{esr_1lrs}) over the joint distribution of ${\{\mathbf{H},\hat{\mathbf{H}\}}}$. Nevertheless, it can be achieved by instead maximizing the average SR (ASR), the short-term expected SR over the conditional CSIT error distribution, for each $\mathbf{\hat{H}}$ of a sufficiently large set of random CSIT realizations. For a given CSIT $\mathbf{\hat{H}}$, the ASR expression is given by
\begin{equation}
    \text{ASR}(\mathbf{P}) = \bar{R}_c + \sum_{g=1}^G \bar{R}_{c,g} + \sum_{k=1}^K \bar{R}_{p,k},
    \label{asr_expression}
\end{equation}
where the individual terms $\bar{R}_{c}\triangleq\min_k\{\mathbb{E}_{\text{H}|\hat{\text{H}}}\{R_{c,k}|\hat{\mathbf{H}}\}\}_{k=1}^K$, $\bar{R}_{c,g}\triangleq\min_{k_g}\{\mathbb{E}_{\text{H}|\hat{\text{H}}}\{R_{c,g,k_g}|\hat{\mathbf{H}}\}\}_{k_g=1}^{K_g}$, $\forall g \in \mathcal{G},$, and  $\bar{R}_{p,k}\triangleq\mathbb{E}_{\text{H}|\hat{\text{H}}}\{R_{k}|\hat{\mathbf{H}}\},$ $\forall k \in \mathcal{K},$ describe the global common, group common, and private average rates (ARs), respectively.

As also proposed in \cite{hamdi}, the sampled average approximation (SAA) method can be used to obtain a deterministic equivalent of the stochastic ASR expression in (\ref{asr_expression}). Thus, for a given CSIT realization $\mathbf{\hat{H}}$ and conditional CSIT error distribution, a set of $M$ i.i.d CSIT error realizations, indexed by the set $\mathcal{M}\triangleq\{1,\dots,M\}$, is generated and it is given by $\Tilde{\mathbb{H}}^{(\textbf{M})}\triangleq\{\Tilde{\mathbf{H}}^{(m)}|\;m \in \mathcal{M}\}$. Assuming the same CSIT error variance $\sigma_e^2$ for all $K$ users, the set of $M$ real CSI realizations related to the CSIT error set $\Tilde{\mathbb{H}}^{(\textbf{M})}$ is in turn given by  
\begin{equation}
    \mathbb{H}^{(\textbf{M})}\triangleq\{\mathbf{H}^{(m)}=\sqrt{1-\sigma_e^2}\hat{\mathbf{H}}+\sigma_e\Tilde{\mathbf{H}}^{(m)}|\;\hat{\mathbf{H}},\;m \in \mathcal{M}\}.
    \label{saa_def}
\end{equation}
\begin{algorithm}[t!]
\DontPrintSemicolon
  \KwInput{$\bar{R}_{c}^{(\textbf{M})}, \{\bar{R}_{c,g}^{(\textbf{M})},K_g\}_{g=1}^G,\{\bar{R}_{k}^{(\textbf{M})}\}_{k=1}^K, K, \{\bar{R}_{k}^{th}\}_{k=1}^K$}
   
  \text{Initialize} $\bar{R}_k^{alloc} = \bar{R}_{k}^{(\textbf{M})}, \forall k \in \mathcal{K} $ \;
  $num\_users\_left = \sum_{k=1}^K 1(\bar{R}_k^{alloc}-\bar{R}_k^{th} < 0)$ \;

\uIf{$num\_users\_left > 0$}{
 \For{$g \leftarrow 1,\dots,G$}{
    $\bar{R}_{c,g}^{avail} = \bar{R}_{c,g}^{(\textbf{M})}$ \;
    \For{$k_g \leftarrow 1,\dots,K_g$}{
        \uIf{$(\bar{R}_{k_g}^{alloc}-\bar{R}_{k_g}^{th}) < 0$}{
            \uIf{$\bar{R}_{c,g}^{avail} > (\bar{R}_{k_g}^{th} - \bar{R}_{k_g}^{alloc})$}{
                $\bar{R}_{c,g}^{avail} = \bar{R}_{c,g}^{avail} - (\bar{R}_{k_g}^{th} - \bar{R}_{k_g}^{alloc})$ \;
                $\bar{R}_{k_g}^{alloc} = \bar{R}_{k_g}^{alloc} + (\bar{R}_{k_g}^{th} - \bar{R}_{k_g}^{alloc})$\;
            }
            \Else{
                $\bar{R}_{k_g}^{alloc} = \bar{R}_{k_g}^{alloc} + \bar{R}_{c,g}^{avail}$\;
                $\bar{R}_{c,g}^{avail} = 0$\;
                \textbf{break}\;
            }
        }
    }
    \uIf{$\bar{R}_{c,g}^{avail} > 0$}{
        $\bar{R}_{k_g}^{alloc} = \bar{R}_{k_g}^{alloc} + \frac{\bar{R}_{c,g}^{avail}}{K_g}\;,\;\forall k_g \in \mathcal{K}_g $\;
    }
 }
 $num\_users\_left = \sum_{k=1}^K 1(\bar{R}_k^{alloc}-\bar{R}_k^{th} < 0)$ \;
 \uIf{$num\_users\_left > 0$}{
    $\bar{R}_{c}^{avail} = \bar{R}_{c}^{(\textbf{M})}$ \;
    \For{$k \leftarrow 1,\dots,K$}{
        \uIf{$(\bar{R}_{k}^{alloc}-\bar{R}_{k}^{th}) < 0$}{
            \uIf{$\bar{R}_{c}^{avail} > (\bar{R}_{k}^{th} - \bar{R}_{k}^{alloc})$}{
                $\bar{R}_{c}^{avail} = \bar{R}_{c}^{avail} - (\bar{R}_{k}^{th} - \bar{R}_{k}^{alloc})$ \;
                $\bar{R}_{k}^{alloc} = \bar{R}_{k}^{alloc} + (\bar{R}_{k}^{th} - \bar{R}_{k}^{alloc})$\;
            }
            \Else{
                $\bar{R}_{k}^{alloc} = \bar{R}_{k}^{alloc} + \bar{R}_{c}^{avail}$\;
                $\bar{R}_{c}^{avail} = 0$\;
                \textbf{break}\;
            }
        }
    }
    \uIf{$\bar{R}_{c}^{avail} > 0$}{
        $\bar{R}_{k}^{alloc} = \bar{R}_{k}^{alloc} + \frac{\bar{R}_{c}^{avail}}{K}\;,\;\forall k \in \mathcal{K} $\;
    }
 }
 \Else{
 $\bar{R}_k^{alloc} = \bar{R}_k^{alloc} + \frac{\bar{R}_{c}^{(\textbf{M})}}{K}\;,\;\forall k \in \mathcal{K}  $\;
 }
}
\Else{
  \For{$g \leftarrow 1,\dots,G$}{
  $\bar{R}_{k_g}^{alloc} = \bar{R}_{k_g}^{alloc} + \frac{\bar{R}_{c,g}^{(\textbf{M})}}{K_g}\;,\;\forall k_g \in \mathcal{K}_g $\;
  }
  $\bar{R}_k^{alloc} = \bar{R}_k^{alloc} + \frac{\bar{R}_{c}^{(\textbf{M})}}{K}\;,\;\forall k \in \mathcal{K}  $\;
  }

  \KwOutput{$\{\bar{R}_k^{alloc}\}_{k=1}^K$}
\caption{Global and Group Common Rates Allocation based on Target QoS Rates}
\label{rate_alloc_alg}
\end{algorithm}
\begin{figure*}[t!]
\begin{minipage}{0.5\linewidth}
\centering
\subfloat[]{\label{case_1_2_1}\includegraphics[scale=.55]{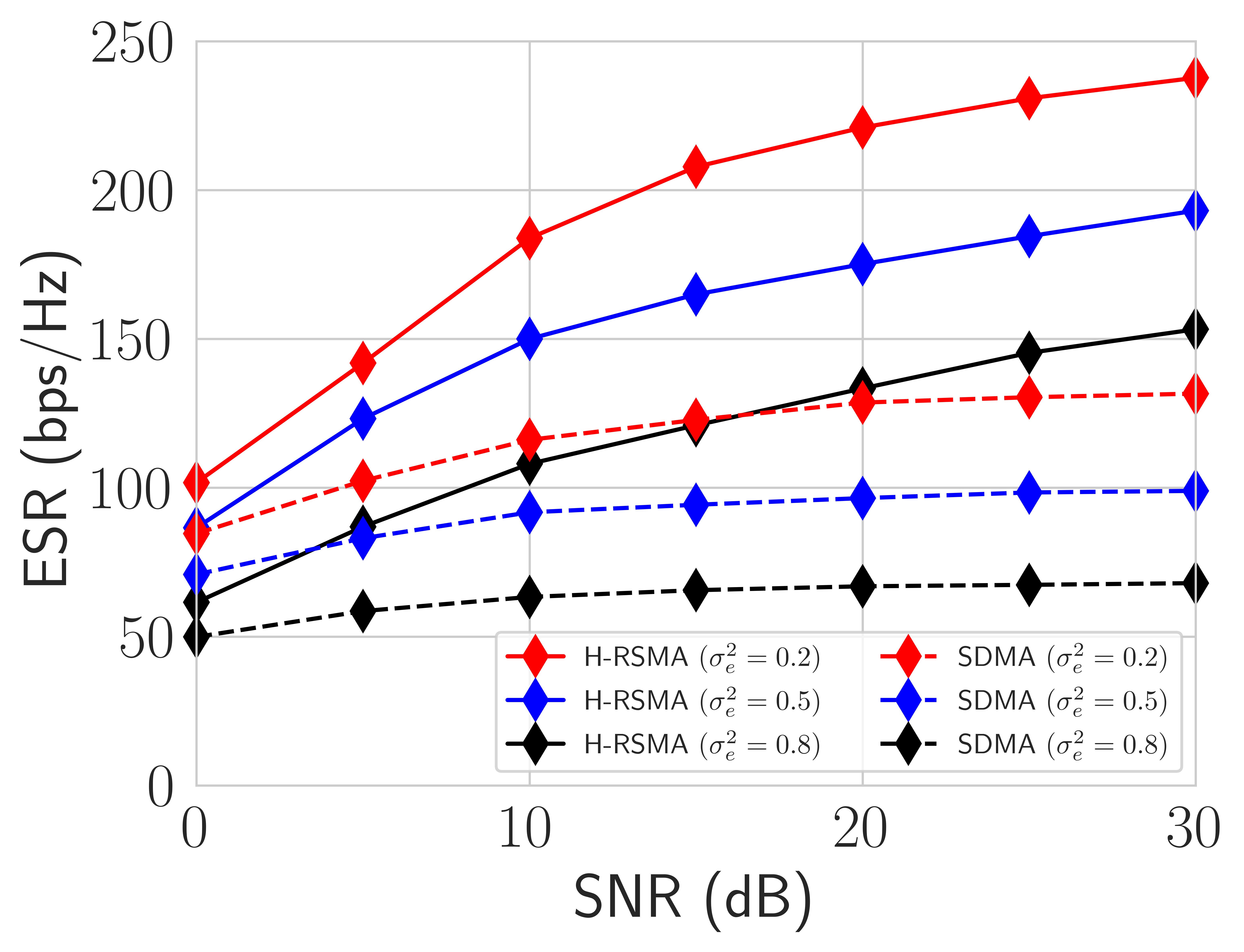}}
\end{minipage}
\begin{minipage}{0.5\linewidth}
\centering
\subfloat[]{\label{case_1_2_2}\includegraphics[scale=.55]{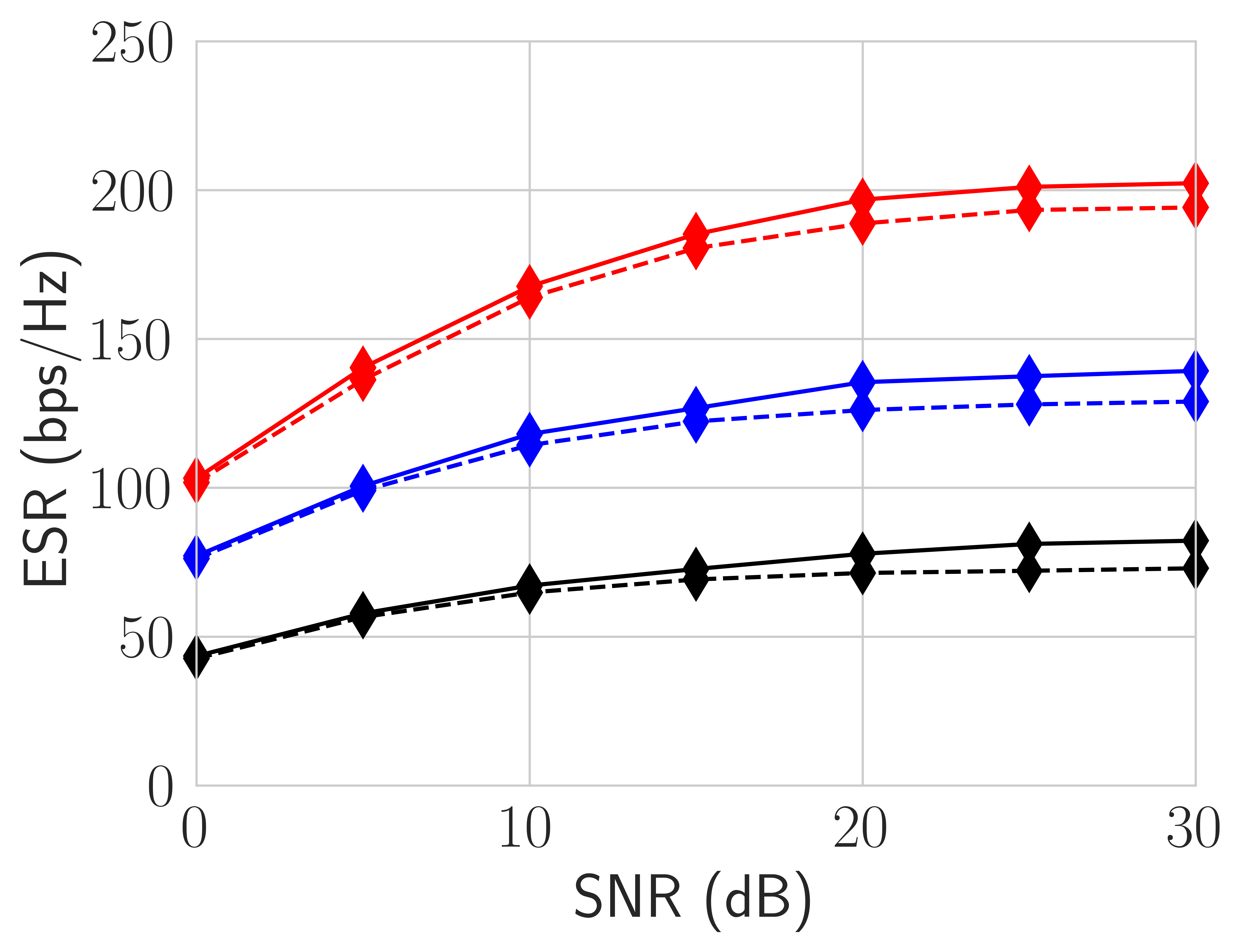}}
\end{minipage}\par\medskip
\caption{ESR vs. SNR ($N_t = 100, K=90$): (a) disjoint user groups ($\Delta_k=\frac{\pi}{36}$), and (b) overlapping user groups ($\Delta_k=\frac{\pi}{8}$).}
\label{fig:case_1_3}
\end{figure*}
According to the strong Law of Large Numbers, the ARs of each stream can be estimated through their sample average functions (SAFs) as $M \rightarrow \infty$. The SAFs are given by $\bar{R}_{c,k}^{(\textbf{M})}\triangleq\frac{1}{M}\sum_{m=1}^MR_{c,k}^{(m)}$, $\bar{R}_{c,g,k_g}^{(\textbf{M})}\triangleq\frac{1}{M}\sum_{m=1}^MR_{c,g,k_g}^{(m)}$, and $\bar{R}_k^{(\textbf{M})}\triangleq\frac{1}{M}\sum_{m=1}^MR_k^{(m)}$, where $R_{c,k}^{(m)}$ and $R_k^{(m)}$ are the achievable rates of the global common and private streams at user-$k$, and $R_{c,g,k_g}^{(m)}$ is the achievable rate at user-$k_g$ for the $m$-th CSI realization $\mathbf{H}^{(m)}$ in the set $\mathbb{H}^{(\textbf{M})}$. The SAA of the ASR maximization problem of H-RSMA can then be defined as follows
\begin{maxi!} |s|[2]
{\mathbf{P}}{\bar{R}_{c}^{(\textbf{M})}+\sum_{g=1}^G\bar{R}_{c,g}^{(\textbf{M})}+\sum_{k=1}^K\bar{R}_{k}^{(\textbf{M})}}{\label{hrs_opt_prob}}{} 
\addConstraint{\bar{R}_{c}^{(\textbf{M})}}{\leq\bar{R}_{c,k}^{(\textbf{M})},\quad}{\forall k\in\mathcal{K}  \label{eq:C1_hrs}}
\addConstraint{\bar{R}_{c,g}^{(\textbf{M})}}{\leq\bar{R}_{c,g,k_g}^{(\textbf{M})},\quad}{\forall k_g\in\mathcal{K}_g\;,\; g\in\mathcal{G} \label{eq:C2_hrs}}
\addConstraint{\Tr(\mathbf{P}\mathbf{P}^H)}{\leq P_t. \label{eq:C3_hrs}}
\addConstraint{\bar{R}_k^{total}}{\geq \bar{R}_k^{th}. \label{eq:C4_hrs}}
\end{maxi!}
where $\bar{R}_{c}^{(\textbf{M})}$ is the SAF of the global common stream; $\bar{R}_{c,g}^{(\textbf{M})}$, the SAF of the $g$-th group common stream; and $\bar{R}_{k}^{(\textbf{M})}$, the SAF of the $k$-th private stream. Also, $\bar{R}_{c,k}^{(\textbf{M})}$ denotes the achievable global common stream rate at user-$k$, $\bar{R}_{c,g,k_g}^{(\textbf{M})}$ is the achievable group common stream rate at user-$k_g$ of group-$g$, $\bar{R}_k^{total}$ denotes the total AR of user-$k$, and $\bar{R}_k^{th}$ denotes the QoS rate constraint for user-$k$. Constraint (\ref{eq:C1_hrs}) indicates that the global common stream rate must be less or equal to the minimum of all achievable global common stream rates for the $K$ users. Similarly, constraint (\ref{eq:C2_hrs}) indicates that the group common stream rate for group-$g$ must be less or equal to the minimum of all achievable group common stream rate among the $K_g$ users in group-g. Finally, constraint (\ref{eq:C4_hrs}) denotes the QoS rate constraint for each of the $K$ users. Based on this optimization problem, the meta-loss function can be defined as follows

\begin{equation}
\begin{split}    
    \mathcal{L}_{meta}(\mathbf{P}) = &- \Big[\min\{\bar{R}_{c,1}^{(\textbf{M})},\dots, \bar{R}_{c,K}^{(\textbf{M})}\} \\ &+\sum_{g=1}^G\min\{\bar{R}_{g,1}^{(\textbf{M})},\dots, \bar{R}_{g,K_g}^{(\textbf{M})}\} +\sum_{k=1}^K\bar{R}_{k}^{(\textbf{M})}\Big] \\
    &- \lambda \sum_{k=1}^K (\bar{R}_k^{alloc}-\bar{R}_k^{th})1(\bar{R}_k^{alloc}-\bar{R}_k^{th} < 0),
\end{split}
\label{asr_maximization_problem}
\end{equation}
where the last term is introduced to replace the QoS rate constraints in (\ref{eq:C4_hrs}). Specifically, $\lambda$ is a regularization parameter to balance the tradeoff between enforcing the QoS rate constraints and prioritizing them over maximizing the ASR, and $\bar{R}_k^{alloc},\forall k \in \mathcal{K}$ is the output of Algorithm \ref{rate_alloc_alg}, which is employed to dynamically allocate the global common and group common rates among the corresponding users based on their target QoS rate. Also, $1(\bar{R}_k^{alloc}-\bar{R}_K^{th} < 0) \in \{0,1\}$ denotes the indicator function used to penalize the meta-loss function based only on the users that do not comply with their corresponding QoS rate constraint.  Although the nature of Algorithm \ref{rate_alloc_alg} is not learning-based but rather hand-crafted and heuristic, numerical results presented in the next subsection reveal that the proposed meta-learning based framework is successful in using it to comply with the QoS rate constraints. Finally, it is highlighted that the complexity of employing Algorithm \ref{meta_alg_1} to solve the ASR maximization problem in (\ref{asr_maximization_problem}) is $\mathcal{O}(2TN_t(K+G+1))$, which is substantially lower than the $\mathcal{O}(T(N_tK)^{3.5})$ complexity of the SAA-WMMSE algorithm proposed in \cite{hamdi}.

\subsection{Numerical results}
\label{subsection_4_c}
Due to the astronomically large complexity of the SAA-WMMSE algorithm, it has not been previously possible to evaluate the optimized performance of H-RSMA in the large scale regime. Therefore, in this section the proposed meta-learning based optimization framework is used to optimize and compare the performance of H-RSMA and SDMA. {blue}{ESR results are obtained by averaging the ASR results over 15 random CSIT realizations. The SAA method is applied considering $M=1000$ random CSIT error realizations. The precoder initialization $\mathbf{P}_0$ is obtained using the SVD-MRT method \cite{lina_dpc}, where the group common precoders $\mathbf{p}_{c,g},\forall g \in \mathcal{G}$ are obtained by considering only the channel vectors of the corresponding users. Also, 70\% of the total power is initially allocated to the global common stream precoder, 25\% is equally distributed among the group common stream precoders, and the remaining 5\% is equally distributed among the private stream precoders. 

Consider a scenario with $N_t=100$ and $K=90$. The transmit antennas are equally spaced in a circular uniform array (UCA) of radius $\lambda D$, where $\lambda$ denotes the wavelength and $D=\frac{0.5}{\sqrt{(1-\cos(2\pi/M))^2+\sin(2\pi/M)^2}}$. The communication users are equally grouped in $G=9$ groups, located in azimuth directions $[\frac{-\pi}{2},\frac{-3\pi}{8},-\frac{\pi}{4},\frac{-\pi}{8},0,\frac{\pi}{8},\frac{\pi}{4},\frac{3\pi}{8},\frac{\pi}{2}]$. Two different settings for the geometrical one-ring scattering model are considered: the first assumes that the user groups are spatially disjoint with scattering rings of angular spread $\Delta_k = \frac{\pi}{36}$, and a second one where the user groups are spatially overlapping with scattering rings of angular spread $\Delta_k = \frac{\pi}{8}$. Also, a QoS rate constraint of $\bar{R}_k^{th}=0.25$ bps/Hz is used for all $K$ users. Finally, it is indicated that the DNN of the base learner contains three hidden layers with 400 neurons each, where the first employs the sigmoid function as activation; and the last two, tanh activation. An Adam optimizer with learning rate $\beta=10^{-3}$.

Fig. \ref{fig:case_1_3} depicts the ESR vs. SNR performance of H-RSMA and SDMA with different degrees of CSIT quality and for the spatially disjoint and overlapping user groups. For the setting with the disjoint user groups, it is observed that the ESR difference between H-RSMA and SDMA is significant across the whole SNR range, and widens as the CSIT quality worsens, reaching a considerable 125\% improvement at 30 dB SNR for $\sigma_e^2=0.8$. In this scenario, the group common streams necessarily play a more crucial role in order to manage the increased intra-group multi-user interference, while the global common stream is relegated to handle inter-group interference only when it cannot be spatially avoided through beamforming. It is also important to indicate that, due to enforcing the QoS rate constraints, the ESR of SDMA rapidly saturates due to operating in an interference-limited state, which is expected as all private stream precoders must be activated. In contrast, the ESR of H-RSMA system does not saturate as the multiple common streams are employed to prioritize scheduling data to the users which have the worst channel conditions while also managing interference to avoid being interference-limited. Thus, it is demonstrated that the ESR gains of H-RSMA over SDMA are not negligible in the large scale regime. On the other hand, for the scenario with overlapping user groups, it can be easily observed that the ESR difference between utilizing H-RSMA and SDMA transmission is minimal, albeit it increases slightly as the CSIT quality worsens reaching an 12\% improvement at 30 dB SNR for $\sigma_e^2=0.8$. To interpret this, it is important to consider that the group common streams are not expected to contribute largely to the ESR due to them being decoded while experiencing the private stream interference of users in the overlapping user groups. Therefore, in order to manage the inter-group multi-user interference, the H-RSMA transmitter resorts to employing the global common stream instead. However, it cannot provide a substantial rate as it needs to be decoded by all 90 users.

\section{Use case 2: Integrated sensing and communication} 
\subsection{System model}
Consider an ISAC transmitter with $N_t$ transmit antennas and imperfect CSIT, that serves $K$ single-antenna communication users, indexed by the set $\mathcal{K} = \{1,\dots,K\}$, while also tracking $N$ radar targets, indexed by the set $\mathcal{N} = \{1,\dots,N\}$, where target-$n$ is located at an azimuth direction $\theta_n$ as depicted in Fig. \ref{fig:isac_model}. In particular, it is considered that the ISAC transmitter employs H-RSMA, as described in Section \ref{hrsma_section}, to serve the communication users, while also employing the same H-RSMA precoded transmit signal $\mathbf{x} \in \mathbb{C}^{N_t \times 1}$ in (\ref{hrsma_tx_signal}) for mono-static MIMO radar tracking, i.e. an additional precoded stream for target tracking is not employed.

\begin{figure}[t!]
		\centering		        
        \includegraphics[width=0.95\columnwidth]{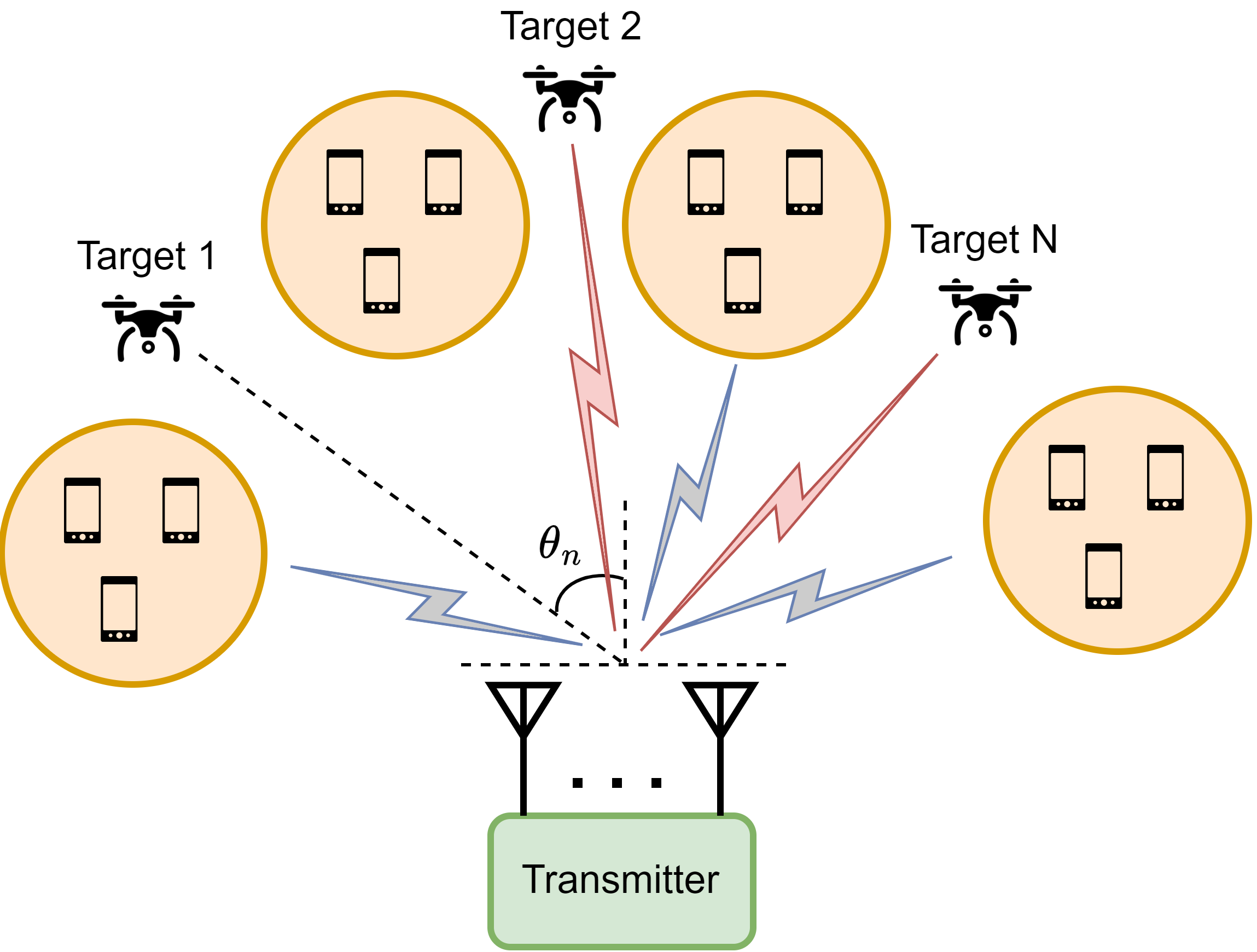}
		\caption{ISAC system model.}
		\label{fig:isac_model}
\end{figure}

\begin{figure*}[t!]
\begin{minipage}{0.5\linewidth}
\centering
\subfloat[]{\label{tradeoff_2}\includegraphics[scale=.55]{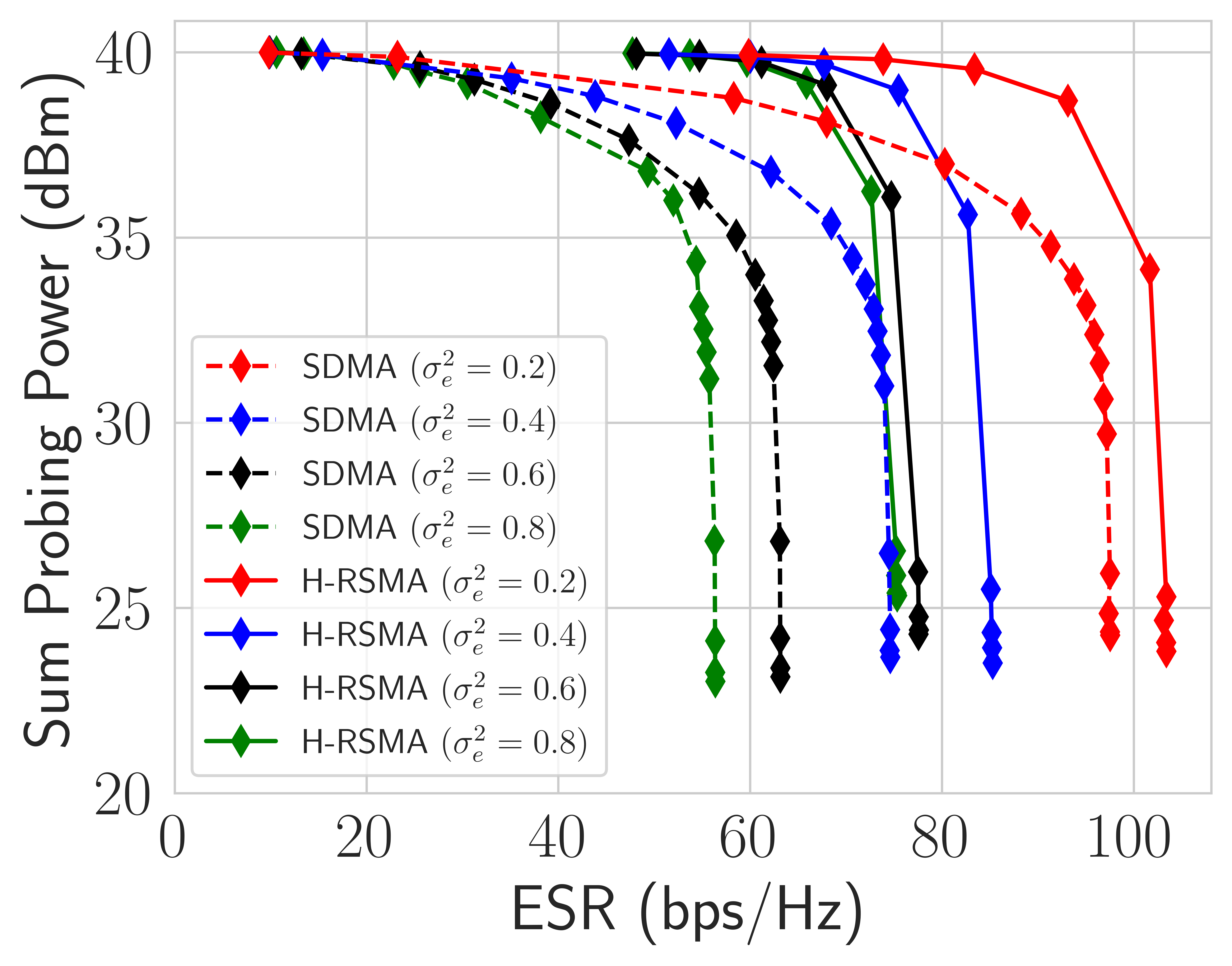}}
\end{minipage}
\begin{minipage}{0.5\linewidth}
\centering
\subfloat[]{\label{tradeoff_1}\includegraphics[scale=.55]{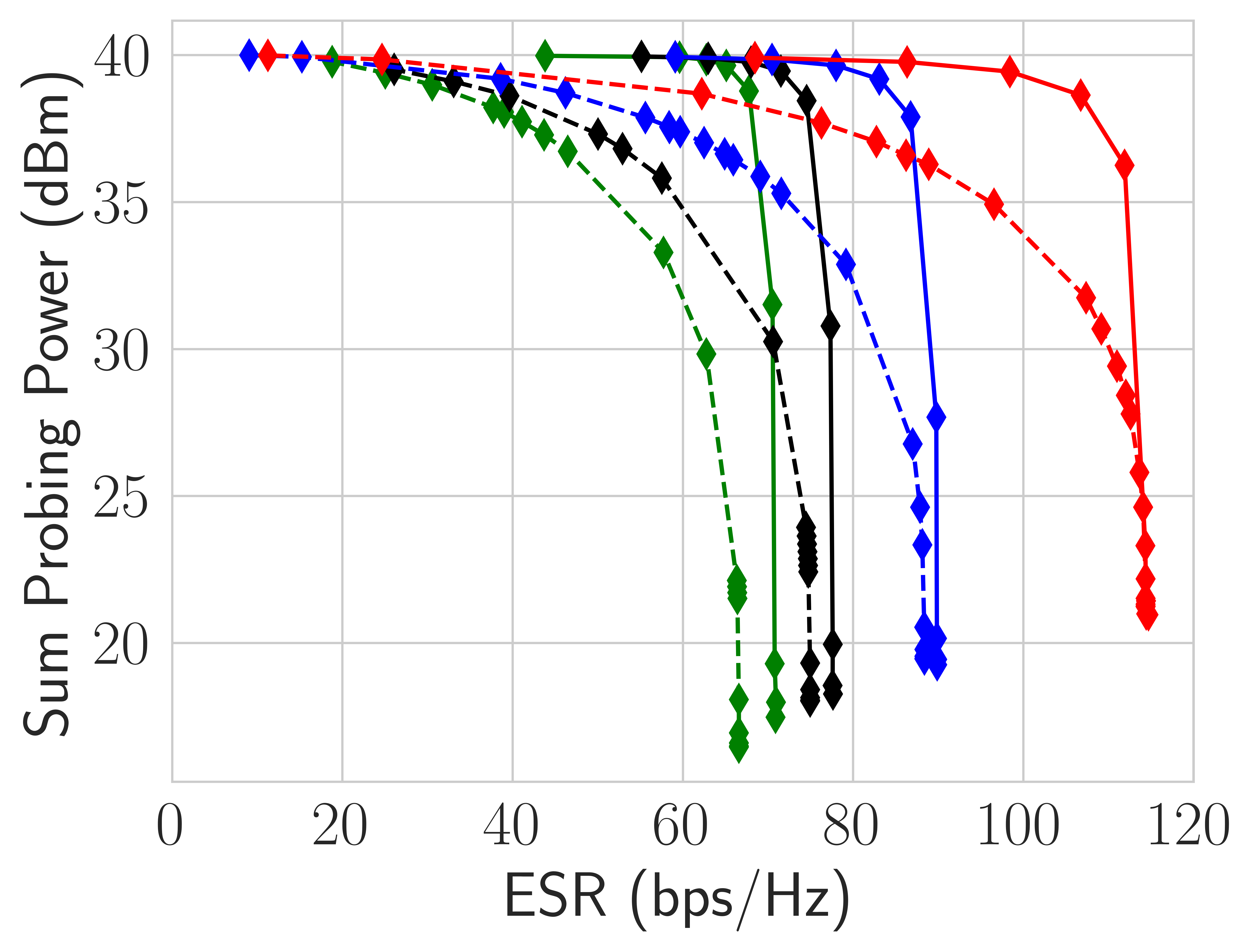}}
\end{minipage}\par\medskip
\caption{ESR vs. sum probing power ($N_t=100, K=40$): (a) disjoint user groups ($\Delta_k = \frac{\pi}{8}$) (b) overlapping user groups ($\Delta_k = \frac{\pi}{3}$)}
\label{fig:tradeoff_fig}
\end{figure*}

\begin{figure*}[t!]
\begin{minipage}{0.5\linewidth}
\centering
\subfloat[]{\label{bpt_1}\includegraphics[scale=.55]{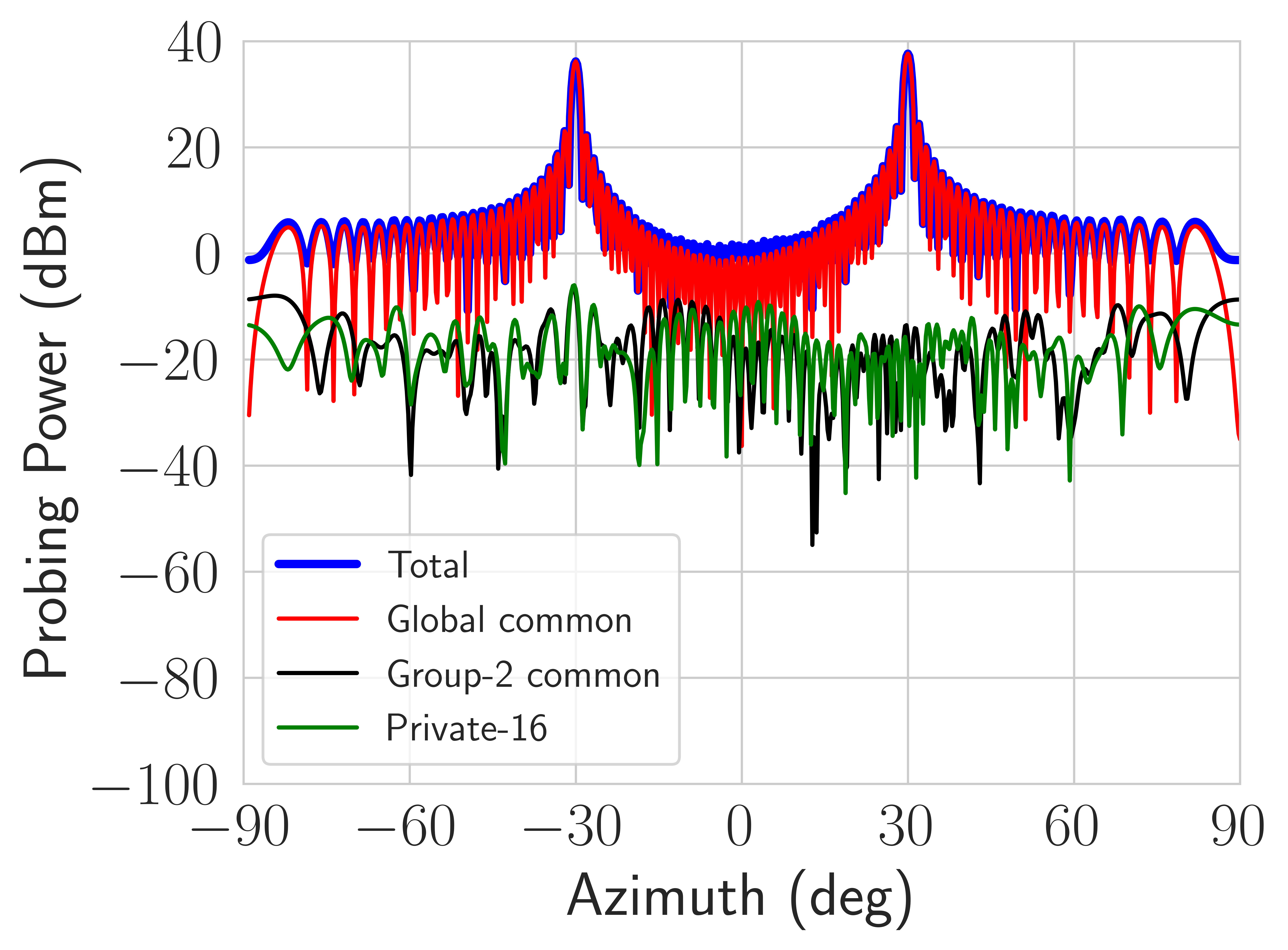}}
\end{minipage}
\begin{minipage}{0.5\linewidth}
\centering
\subfloat[]{\label{bpt_2}\includegraphics[scale=.55]{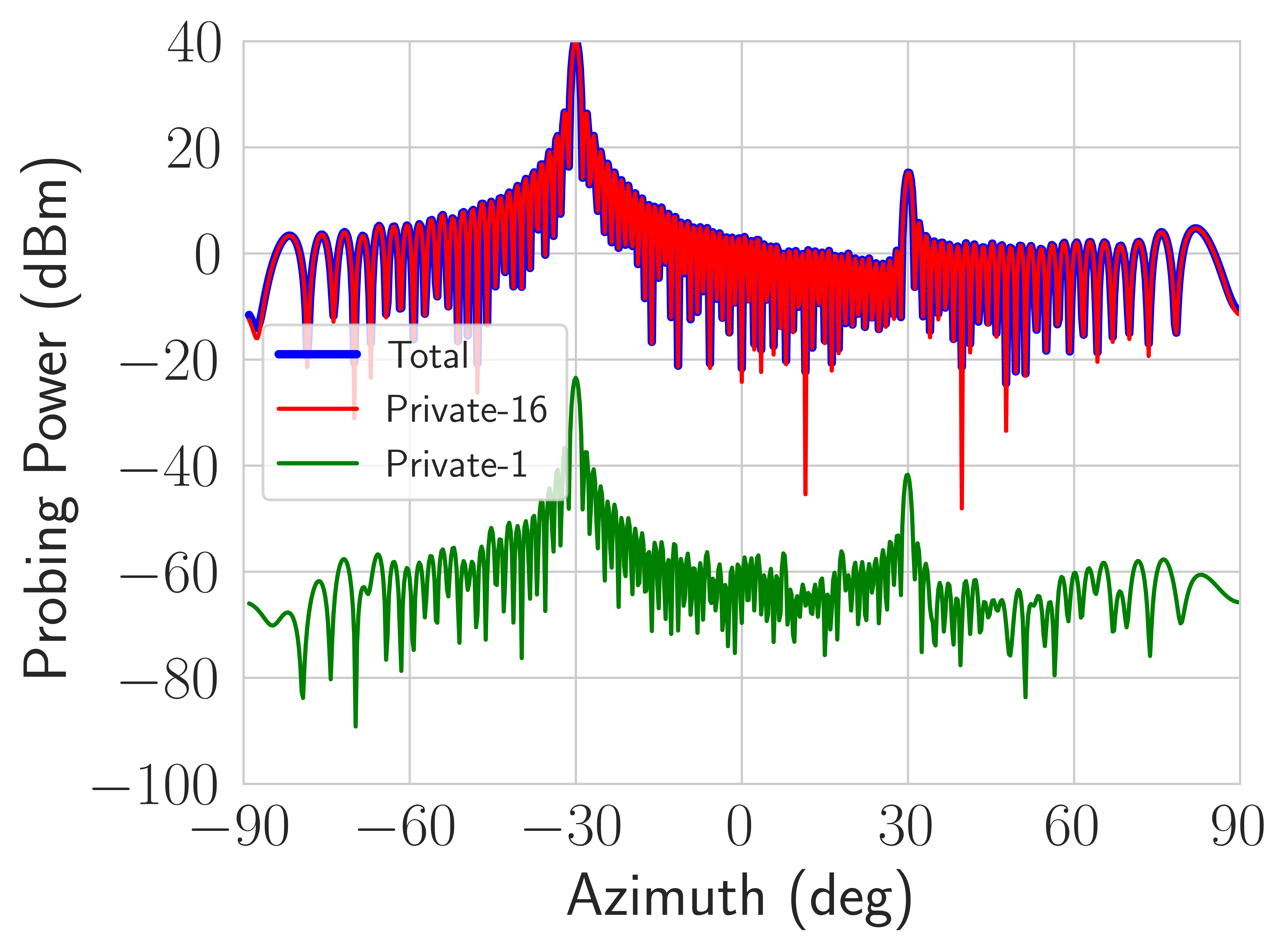}}
\end{minipage}\par\medskip
\caption{Generated beampatterns for the disjoint user groups configuration ($N_t=100, K=40, \Delta_k = \frac{\pi}{8}, \sigma_e^2 = 0.8$): (a) H-RSMA ISAC system. (b) SDMA ISAC system.}
\label{fig:bpt_fig}
\end{figure*}

\subsection{Problem formulation}
As discussed in \cite{rafael_ISAC_2}, the sensing capabilities of a MIMO radar directly depend on appropriately designing its covariance matrix $\mathbf{R}_{\mathbf{x}}=\mathbb{E}\{\mathbf{x}\mathbf{x}^H\}$. Given that it is also assumed that $\mathbb{E}\{\mathbf{s}\mathbf{s}^H\}=\mathbf{I}_{(K+G+1)}$, the covariance matrix is equivalent to $\mathbf{R}_{\mathbf{x}}=\mathbf{P}\mathbf{P}^H$. Thus, its optimization is effectively reduced to optimizing the precoder matrix $\mathbf{P}$ to maximize a relevant radar metric. Although various radar metrics can be used to characterize the radar sensing performance of a MIMO radar, the probing power at the direction of the radar target is one the simplest, yet most decisive factors \cite{probing_pow_ref}. Therefore, the ISAC precoder optimization problem to jointly maximize the ASR and the sum probing power in the directions of the radar targets can be expressed as follows
\begin{maxi} |s|[2]
{\mathbf{P}}{\underbrace{\bar{R}_c^{(\textbf{M})} + \sum_{g=1}^G \bar{R}_{c,g}^{(\textbf{M})} + \sum_{k=1}^K \bar{R}_k^{(\textbf{M})}}_{\text{ASR}} + \lambda \underbrace{\sum_{n=1}^N \mathbf{a}_t^H(\theta_n)(\mathbf{P}\mathbf{P}^H)\mathbf{a}_t(\theta_n)}_{\text{Probing Power}}}{\label{hrs_ISAC_opt_prob}}{} 
\addConstraint{}{(\text{\ref{eq:C1_hrs}}), (\text{\ref{eq:C2_hrs}}), (\text{\ref{eq:C3_hrs}}),}
\end{maxi}
where $\mathbf{a}_t(\theta_n) = [1, e^{j2\pi\delta\sin(\theta_n)},\dots,e^{j2\pi(N_t-1)\delta\sin(\theta_n)}]^T \in \mathbb{C}^{N_t \times 1}$ is the transmit antenna array steering vector at direction $\theta_n$ and $\delta$ is the normalized distance between antennas in units of wavelengths. Finally, $\lambda$ is a regularization parameter used to manage the trade-off between maximizing the system ASR or maximizing the probing power in the direction of the radar targets. Therefore, the definition of the meta-loss function is given as follows
\begin{equation}
\begin{split}    
    \mathcal{L}_{meta}(\mathbf{P}) = &- \Big[\min\{\bar{R}_{c,1}^{(\textbf{M})},\dots, \bar{R}_{c,K}^{(\textbf{M})}\} \\ &+\sum_{g=1}^G\min\{\bar{R}_{g,1}^{(\textbf{M})},\dots, \bar{R}_{g,K_g}^{(\textbf{M})}\}+\sum_{k=1}^K\bar{R}_{k}^{(\textbf{M})}\Big] \\
    &- \lambda \sum_{n=1}^N \mathbf{a}_t^H(\theta_n)(\mathbf{P}\mathbf{P}^H)\mathbf{a}_t(\theta_n).
\end{split}
\end{equation}
As the meta-loss function only depends on the precoder matrix optimization, the single-variable meta-learning based optimization algorithm described in Algorithm \ref{meta_alg_1} can also be employed to perform ISAC precoder optimization with a complexity of $\mathcal{O}(2TN_t(K+G+1))$, which is significantly lower than the $\mathcal{O}(T(N_tK)^{3.5})$ complexity of the ADMM optimization algorithm based on the SAA-WMMSE algorithm proposed in \cite{rafael_ISAC_2} for RSMA ISAC with imperfect CSIT.

\subsection{Numerical results}
In this section, the proposed meta-learning based optimization framework is employed to optimize an H-RSMA ISAC system and assess its performance gains over an SDMA ISAC system with meta-learned precoders in terms of the tradeoff between ESR and the sum probing power at the direction of the radar targets. Additionally, it is of special interest to analyze the contribution of the group common stream precoders in jointly maximizing the ASR and the sum probing power as the works in \cite{rafael_ISAC_2, probing_pow_ref} have demonstrated that the (global) common stream of 1-Layer RSMA is the key element to achieve this by creating a highly directional beampattern. A setting with $N_t=100$ and $K=40$ is considered. The transmit antennas are arranged equally spaced in a UCA manner. are The communication users are equally grouped in $G=4$ groups, located in azimuth directions $[\frac{-\pi}{2},\frac{-\pi}{6},\frac{\pi}{6},\frac{\pi}{2}]$. Two different settings for the geometrical one-ring scattering model are considered: the first assumes that the user groups are spatially disjoint with scattering rings of angular spread $\Delta_k = \frac{\pi}{8}$, and a second one where the user groups are spatially overlapping with scattering rings of angular spread $\Delta_k = \frac{\pi}{3}$. Also, no QoS rate constraints are employed. It is further considered that the ISAC transmitter tracks $N=2$ radar targets at the $[\frac{-\pi}{6}, \frac{\pi}{6}]$ azimuth direction. It is also considered that the regularization parameter varies in the range $\lambda \in  \{10^{-5},10^{-4.5},\dots,10^{-1}\}$ to switch the priority from communications to radar sensing. Finally, the meta-learning based optimization algorithm is run for $T=7500$ iterations. Finally, it is indicated that the same parameters of the DNN of the base learner used in Subsection \ref{subsection_4_c} are used.

Fig. \ref{fig:tradeoff_fig} portrays the ESR vs. sum probing power trade-off of the H-RSMA ISAC  and SDMA ISAC systems for different CSIT quality degrees. It can be immediately observed that the ESR of the SDMA ISAC system rapidly drops as the sum probing power is prioritized in the meta-learning based optimization process. Specifically for the disjoint user groups deployment, regardless of the CSIT quality, the ESR of the SDMA ISAC system reaches an ESR around 11 bps/Hz when radar sensing is fully prioritized by using a regularization parameter value of $\lambda = 10^{-1}$. In contrast, the ESR of the H-RSMA ISAC system achieves an ESR of 48 bps/Hz when $\sigma_e^2 = 0.8$ and radar is fully prioritized. Thus, this reveals that the common streams of the H-RSMA ISAC system play a crucial role in offering an astonishing 336\% ESR improvement over the SDMA ISAC system when fully prioritizing the maximization of the probing power for radar functionalities. This trend is also observed when a deployment with overlapping user groups is considered. For $\lambda = 10^{-1}$ and $\sigma_e^2 = 0.8$, the SDMA ISAC system only achieves an ESR of 8 bps/Hz while the H-RSMA ISAC system achieves an ESR of 42 bps/Hz (a 425\% ESR improvement). It can also be noticed that the H-RSMA ISAC successfully mitigates the ESR loss when switching from prioritizing communications to radar. For instance, in the overlapping user groups scenario with $\sigma_e^2 = 0.2$, the ESR of the H-RSMA ISAC drops 40\% whereas the ESR of the SDMA ISAC drops 90\%.

To illustrate the crucial roles of the common stream precoders, consider the beampattern plots shown in Fig. \ref{fig:bpt_fig} for the disjoint user group deployment with $\lambda=10^{-1}$, where it is confirmed that the H-RSMA ISAC system primarily employs the global common stream to maximize the probing power by generating highly directional beampattern lobes in the directions of the radar targets. In contrast, the SDMA ISAC system converges to a solution in which only one private stream is employed to generate the directional radar beampattern, albeit a significant disparity in the probing power at the different radar target directions is also observed. Thus, it is revealed that, although the group common stream precoders are not primarily employed to maximize the sum probing power, they still play an important role in managing the intra-group interference to maximize the ASR.

\section{Use case 3: Beyond-diagonal reconfigurable intelligent surfaces}

\subsection{System model}
Consider a RIS-aided communication system where a transmitter, equipped with $N_t$ antennas, communicates with $K$ users, indexed by the set $\mathcal{K}=\{1,\dots,K\}$, with the assistance of an RIS equipped $B$ scattering elements, indexed by the set $\mathcal{B}=\{1,\dots,B\}$. This RIS aids in improving the channel conditions between the transmitter and the $K$ communication users due to a lack of direct link between them, as depicted in Fig. \ref{fig:bdris_model}. The transmitter schedules the messages for the $K$ users and encodes them and modulates them into $K$ independent data streams denoted by $\mathbf{s} = [s_1,\dots,s_K]^T \in \mathbb{C}^{K \times 1}$. It then uses the precoder matrix $\mathbf{P} \in \mathbb{C}^{N_t \times K}$ to generate the transmit signal $\mathbf{x} = \mathbf{P}\mathbf{s} \in \mathbb{C}^{N_t \times 1}$. At the receiver of user-$k$, the received signal is given by 
\begin{equation}
    y_k = \mathbf{h}_k^H\mathbf{\Phi}\mathbf{G}\mathbf{P}\mathbf{s}+n_k,
    \label{bd_ris_rx_signal_1}
\end{equation}
where $\mathbf{h}_k\in\mathbb{C}^{M\times 1}$ is the channel vector between the RIS and user-$k$, $\mathbf{G}\in \mathbb{C}^{B\times N_t}$ denotes the channel matrix between the transmitter and the RIS, and $n_k\;\mathtt{\sim}\; \mathcal{CN}(0,\sigma_{n,k}^2)$ is the AWGN at the output of the receiver antenna. Additionally, the matrix $\mathbf{\Phi} \in \mathbb{C}^{B \times B}$ represents the scattering matrix of the $B-$port RIS, in which the element $[\mathbf{\Phi}]_{i,j}$ describes the amplitude and phase change of an electromagnetic wave that impinges on element $i$ and is then reflected through element $j$. The SINR of decoding its own stream at user-$k$ is

\begin{equation}
    \gamma_k = \frac{|\mathbf{h}_k^H\mathbf{\Phi}\mathbf{G}\mathbf{p}_k|^2}{\sum_{j=1,j\neq k}^K|\mathbf{h}_k^H\mathbf{\Phi}\mathbf{G}\mathbf{p}_j|^2+\sigma_{n,k}^2}.
\end{equation}
Therefore, assuming ideal Gaussian signalling and infinite blocklength codes, the achievable rate of user-$k$ is given by $R_k=\log_2(1+\gamma_k)$. 

\begin{figure}[t!]
		\centering		        
        \includegraphics[width=0.95\columnwidth]{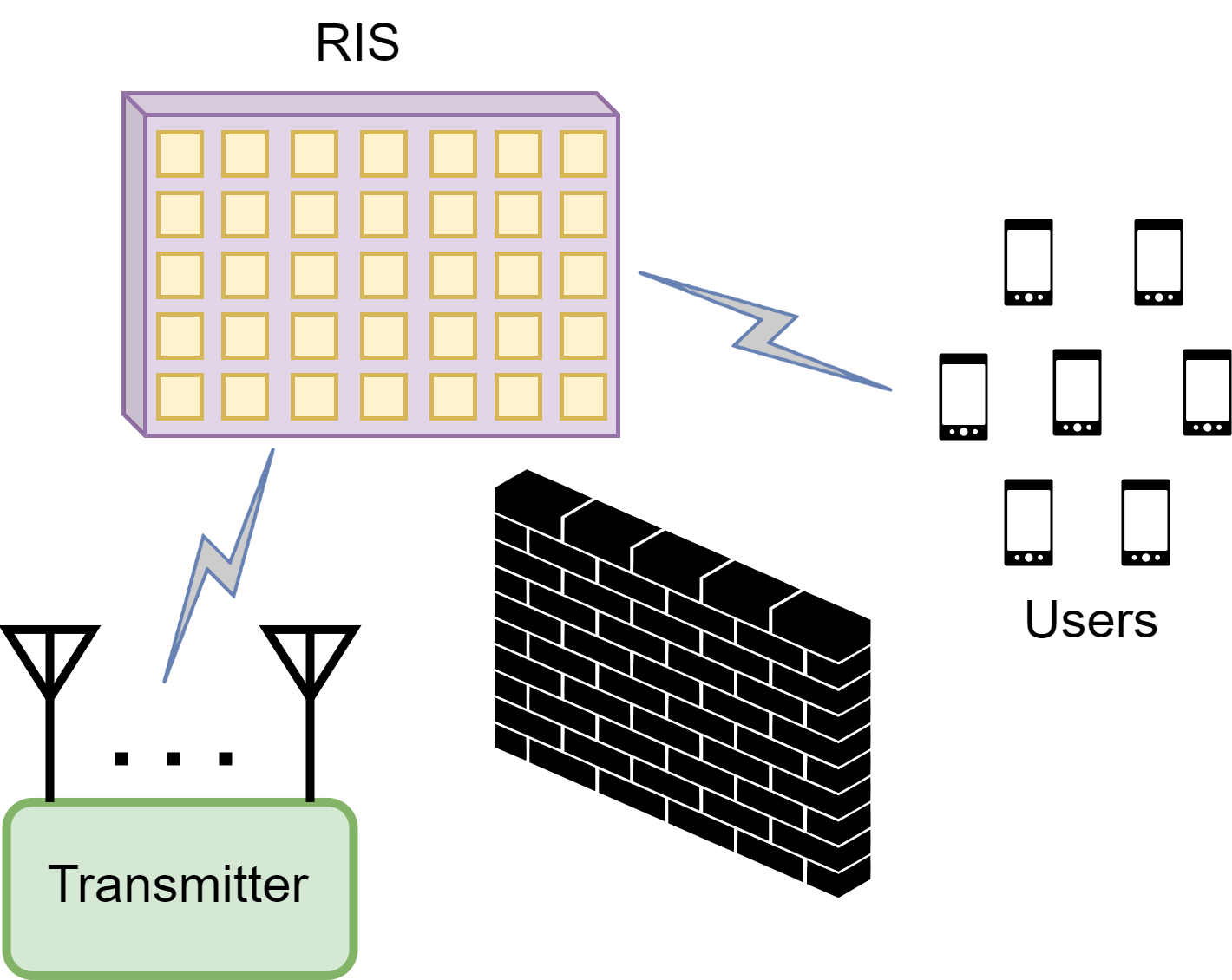}
		\caption{RIS-aided communication system model.}
		\label{fig:bdris_model}
\end{figure}
\subsection{Problem formulation}
Differently from the previous two use cases, the optimization of a RIS-aided communication system is a dual-variable problem due to the presence of the independent variables $\mathbf{P}$ and $\mathbf{\Phi}$. Also, the scattering matrix $\mathbf{\Phi}$ has different structures depending on the interconnections between its scattering elements. Thus, if a conventional RIS structure is considered, the scattering matrix $\mathbf{\Phi}$ is diagonal and given by
\begin{equation}
    \mathbf{\Phi} = \text{diag}(\phi_1,\dots,\phi_B),
\end{equation}
where, if it is assumed that the RIS is lossless, the following constraints must be satisfied
\begin{equation}
    |\phi_b|^2 = 1\;,\;\forall\; b \in \mathcal{B}.
    \label{const_1}
\end{equation}
In turn, if a fully-connected BD-RIS structure is assumed instead, all RIS elements are interconnected with each other and the scattering matrix $\mathbf{\Phi}$ is non-diagonal and satisfies the constraint
\begin{equation}
    \mathbf{\Phi}^H\mathbf{\Phi} = \mathbf{I}_B.
    \label{const_2}
\end{equation}
Thus, the BD-RIS structure can provide the best performance, due to the more general constraint, at the expense of having to jointly optimize the $B^2$ elements in the scattering matrix $\mathbf{\Phi}$ \cite{hongyu_bd_ris}. The SR maximization problem of the RIS-aided system is then given by

\begin{maxi!} |s|[2]
{\mathbf{P},\mathbf{\Phi}}{\sum_{k=1}^K{R}_{k}}{\label{bdris_opt_prob}}{} 
\addConstraint{\Tr(\mathbf{P}\mathbf{P}^H)}{\leq P_t, \label{eq:C1_bdris}}
\addConstraint{f_1(\mathbf{\Phi})}{ \label{eq:C2_bdris}}
\end{maxi!}
where the term $f_1(\mathbf{\Phi})$ is given by 
\begin{equation}
    f_1(\mathbf{\Phi}) = \begin{cases}
        (\ref{const_1}), \text{ if RIS} \\
        (\ref{const_2}), \text{ if BD-RIS}
    \end{cases}.
\end{equation}

\begin{algorithm}[t!]
\DontPrintSemicolon
  \KwInput{$N_t, K, {\mathbf{H}}, T, \beta^1, \beta^2.$}
  
  \text{Initialize} $\mathbf{P}_0$, $\mathbf{\Phi}_0$, $\bm{\theta}_0$

   $\mathcal{L}^{*}_{meta} = \mathcal{L}_{meta}(\mathbf{P}_0, \mathbf{\Phi}_0)$
  
  \For{$t \leftarrow 0,1,\dots,T-1$}
  {$\Delta\mathbf{P}_t = \text{G}_{\bm{\theta}^1_t}^1(\nabla_{\mathbf{P}_0}\mathcal{L}_{meta}(\mathbf{P}_0,\mathbf{\Phi}_0))$\;
  $\mathbf{P}_{t+1} = \mathbf{P}_0 + \Delta\mathbf{P}_t$ \;
  $\mathbf{P}_{t+1} = \Omega(\mathbf{P}_{t+1})$\;
  $\Delta\mathbf{\Phi}_t = \text{G}_{\bm{\theta}^2_t}^2(\nabla_{\mathbf{\Phi}_0}\mathcal{L}_{meta}(\mathbf{P}_0,\mathbf{\Phi}_0))$\;
  $\mathbf{\Phi}_{t+1} = \mathbf{\Phi}_0 + \Delta\mathbf{\Phi}_t$ \;
  \uIf{$\mathcal{L}_{meta}(\mathbf{P}_{t+1},\mathbf{\Phi}_{t+1})<\mathcal{L}^{*}_{meta}$}{
    $\mathcal{L}^{*}_{meta} = \mathcal{L}_{meta}(\mathbf{P}_{t+1},\mathbf{\Phi}_{t+1})$ \;
    $\mathbf{P}^{*} = \mathbf{P}_{t+1}$ \;
    $\mathbf{\Phi}^{*} = \mathbf{\Phi}_{t+1}$ \;
  }
  $\Delta\bm{\theta}_t^1 = \beta^1 \cdot \text{Adam}(\nabla_{\bm{\theta}^1_{t}}\mathcal{L}_{meta}(\mathbf{P}_{t+1},\mathbf{\Phi}_{t+1}))$\;
  $\bm{\theta}^1_{t+1} = \bm{\theta}^1_{t} +\Delta\bm{\theta}^1_t$ \;
  $\Delta\bm{\theta}_t^2 = \beta^2 \cdot \text{Adam}(\nabla_{\bm{\theta}^2_{t}}\mathcal{L}_{meta}(\mathbf{P}_{t+1},\mathbf{\Phi}_{t+1}))$\;
  $\bm{\theta}^2_{t+1} = \bm{\theta}^2_{t} +\Delta\bm{\theta}^2_t$ \;
  }
  \KwOutput{$\mathcal{L}^{*}_{meta},\mathbf{P}^{*},\mathbf{\Phi}^{*}$}
\caption{Dual-Variable Meta-Learning Based Optimization Algorithm}
\label{meta_alg_2}
\end{algorithm}

Due to the dual-variable nature of (\ref{bdris_opt_prob}), a dual-variable meta-learning based optimization algorithm can be employed to solve it, as summarized in Algorithm \ref{meta_alg_2}. The meta-loss function is expressed as follows
\begin{equation}
\begin{split}    
    \mathcal{L}_{meta}(\mathbf{P},\mathbf{\Phi}) = &- \sum_{k=1}^K{R}_{k} + \lambda f_2(\mathbf{\Phi}),
\end{split}
\end{equation}
where the term $f_2(\mathbf{\Phi})$ is given by 
\begin{equation}
    f_2(\mathbf{\Phi}) = \begin{cases}
        ||\bm{\phi}-\mathbf{1}_B||_2^2, \text{ if RIS} \\
        ||\mathbf{\Phi}^H\mathbf{\Phi}-\mathbf{I}_B||_F^2, \text{ if BD-RIS}
    \end{cases},
\end{equation}
where $\bm{\phi} = \text{Diag}(\mathbf{\Phi}) =  e^{j{\bm{\omega}}}\in\mathbb{C}^{B\times 1}$, ${\bm{\omega}} = [\omega_1,\dots,\omega_B]^T \in \mathbb{R}_+^{B\times 1}$ and $\omega_b \in [0,2\pi), \forall b \in \mathcal{B}$. In order to reduce the complexity and training time of $\text{G}_{\bm{\theta}^2_t}^2(.)$, the variable $\bm{\omega}$ can be alternatively considered as its input for RIS optimization, whereas for BD-RIS optimization, and if it is further considered that $\mathbf{\Phi}$ is reciprocal so that $\mathbf{\Phi}^T = \mathbf{\Phi}$ \cite{shanpu_ris}, then $\mathbf{\Phi}$ can be first formatted as an upper triangular matrix of $\frac{B(B+1)}{2}$ non-zero complex elements. Finally, given that both DNNs $\text{G}_{\bm{\theta}^1_t}^1(.)$ and $\text{G}_{\bm{\theta}^2_t}^2(.)$ in Algorithm \ref{meta_alg_2} can be trained independently and in parallel, it is indicated that the complexity of RIS meta-learning based optimization is $\mathcal{O}(T[\max\{2N_tK, B\}])$, and the complexity of BD-RIS meta-learning based optimization is $\mathcal{O}(T[\max\{2N_tK, B(B+1)\}])$. In contrast, the conventional optimization algorithm proposed in \cite{hongyu_2} has a much larger complexity of $\mathcal{O}(T(K^2B^2+T_1KN_t^3+T_2T_3B))$ for RIS optimization, and of $\mathcal{O}(T(K^2B^2+T_1KN_t^3+T_2T_3B^3))$ for BD-RIS optimization ($T_1, T_2$ and $T_3$ are the number of iterations of inner optimization algorithms).

\subsection{Numerical results}

\begin{figure*}[t]
\centering
\begin{minipage}{\columnwidth}
		\centering		        
        \includegraphics[width=\columnwidth]{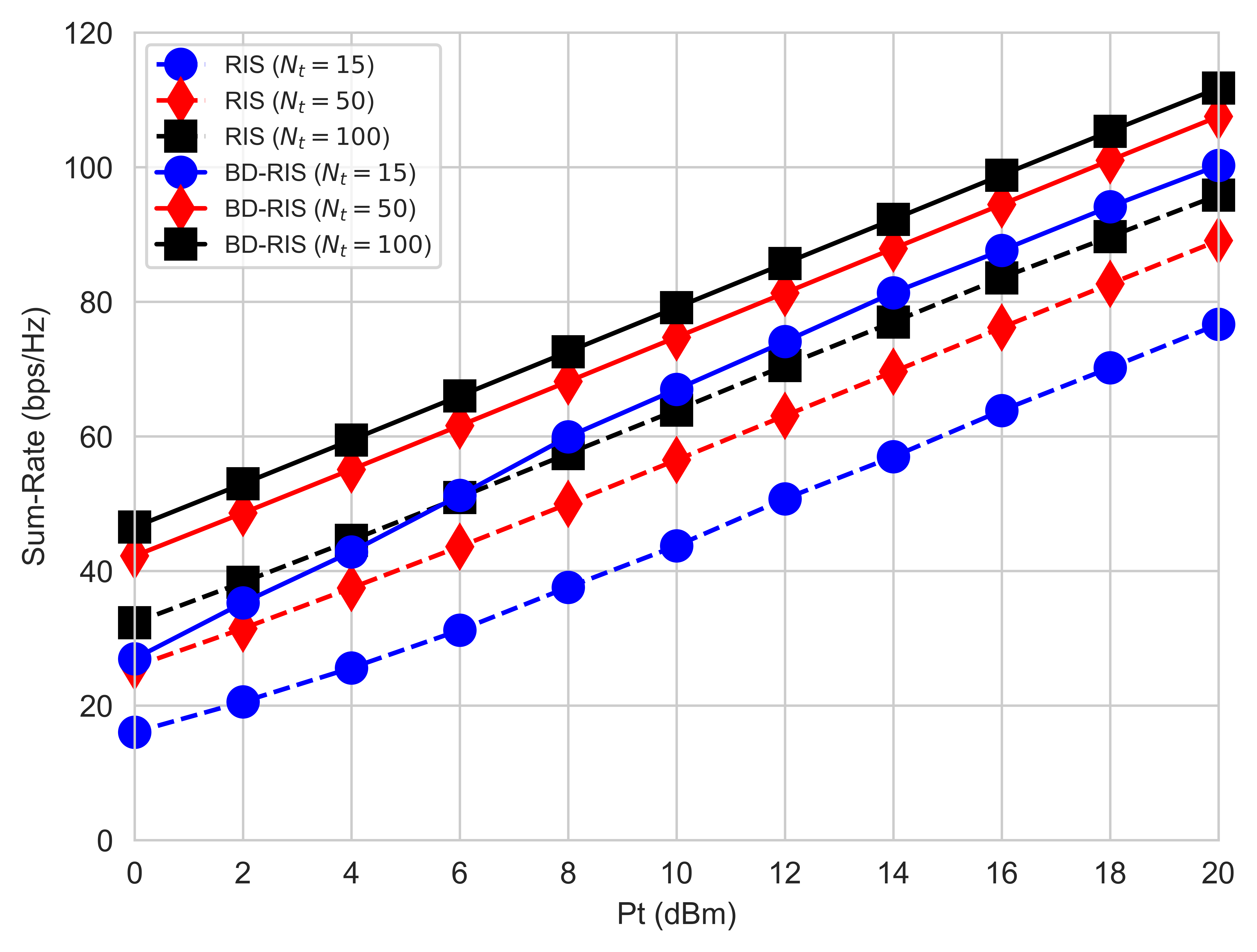}    
		\caption{ESR vs. SNR ($K=10, B=200$): varying $N_t$.}
		\label{fig:bdris_nt}
\end{minipage}
\begin{minipage}{\columnwidth}
		\centering		        
        \includegraphics[width=\columnwidth]{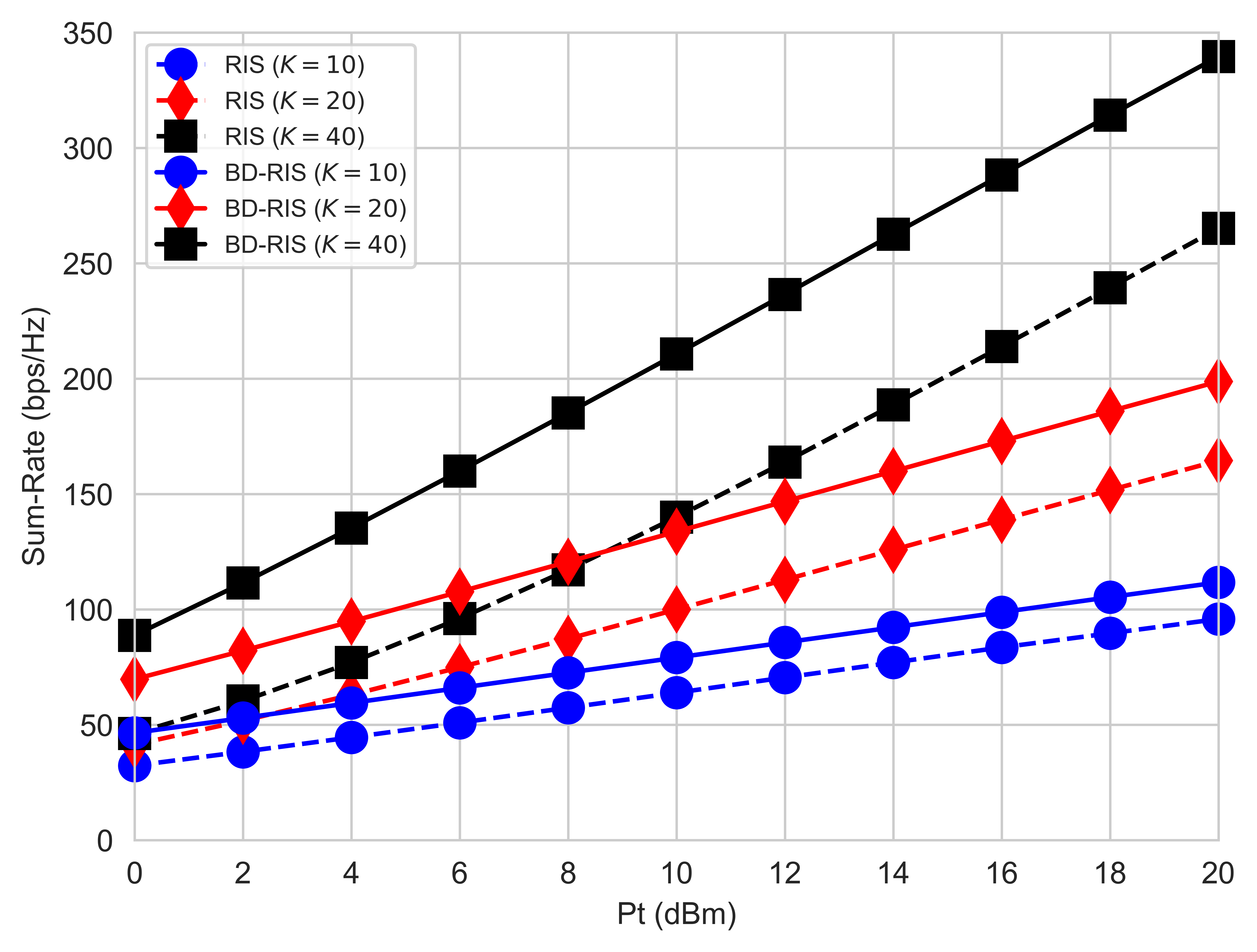}    
		\caption{ESR vs. SNR ($N_t=100, B=200$): varying $K$.}
		\label{fig:bdris_k}
\end{minipage}
\end{figure*}
Due to the high complexity of the conventional optimization algorithm for RIS and BD-RIS, the effect of increasing $N_t, K$ and $B$ in the large scale regime has not been previously evaluated. Therefore, in this section the performance of the proposed meta-learning based optimization algorithm in solving the problem in (\ref{bdris_opt_prob}) is assessed in terms of the achieved SR using different values for $N_t, K$ and $B$. All numerical results are obtained by averaging the results over 50 random channel realizations. Also, it is considered that the channels $\mathbf{G}=\sqrt{\xi_\text{BR}}\Breve{\mathbf{G}}$ and $\mathbf{h}_k=\sqrt{\xi_\text{RU}}\Breve{\mathbf{h}}_k, \forall k \in \mathcal{K}$ \cite{hongyu_bd_ris} are composed by a small-scale fading component ($\Breve{\mathbf{G}}, \Breve{\mathbf{h}}_k$) and a large scale one ($\xi_\text{BR}, \xi_\text{RU}$). The former is considered as Rayleigh fading while the latter is expressed as $\xi_{i} = \xi_0(d_i/d_0)^{-\epsilon_i}, \forall i \in \{\text{BR}, \text{RU}\}$, where $\xi_0$ denotes the signal attenuation at the reference distance $d_0$, $d_\text{BR}$ and $d_\text{RU}$ denote, respectively, the distance between the transmitter and the RIS, and the distance between the RIS and the communication users, and $\epsilon_i,  \forall i \in \{\text{BR}, \text{RU}\}$ is the path-loss exponent. In all simulations, it is considered that $\xi_0 = -30$ dB, $d_0 = 1$ m for all channels, $\epsilon_{\text{BR}}=\epsilon_{\text{RU}}=2$, and $\sigma_{n,k}^2 = -80$ dBm $, \forall k \in \mathcal{K}$. It is also considered that the distance between the transmitter and the RIS is $d_{\text{BR}}=50$ m, and the distance between the RIS and the $K$ users is $d_{\text{RU}} = 2.5$ m. Finally, it is indicated that the precoder initialization $\mathbf{P}_0$ is obtained using the MRT method, and the scattering matrix initialization $\mathbf{\Phi}_0$ for the BD-RIS optimization is obtained as the meta-learned output of the conventional RIS optimization problem with $T=1000$ iterations. The two DNNs in the base learner each contain four hidden layers using the same activation functions described in Subsection \ref{subsection_4_c} in the first three hidden layers while the last one does not employ any activation. The Adam optimizer of the DNN that outputs $\Delta\mathbf{P}$ is $\beta_1=10^{-3}$; and for the DNN that outputs $\Delta\mathbf{\Phi}$, $\beta_2=10^{-4}$. The meta-learning based optimization algorithm is run for $T=25000$ iterations.

The effect of varying $N_t$ is first examined, and results are plotted in Fig. \ref{fig:bdris_nt}. It is mainly observed that, as $Pt$ is increased, the slope of all curves is the same as the number of degrees of freedom (DoFs) is maintained. Instead, its effect is to improve the wireless channel conditions between the transmitter and the RIS through the effects of channel hardening. Thus, it can be observed from both BD-RIS and conventional RIS curves that the SR improvement as $N_t$ increases becomes smaller due to the $\mathbf{G}$ behaving more deterministically. Finally, it is also worth indicating that the fully connected BD-RIS structure significantly outperforms conventional RIS in high SNR, even in the extreme case in which the BD-RIS transmitter features $N_t=15$ antennas; and the RIS transmitter, $N_t=100$ antennas. Therefore, this hints at the potential of BD-RIS to massively improve the SR in scenarios in which increasing the complexity of the transmitter is unfeasible.

The effect of varying $K$ is investigated next, and results are plotted in Fig. \ref{fig:bdris_k}. It is immediately noticed that increasing the number of users derives in an increase in the number of DoFs, which is reasonable due to the larger number of scheduled users. However, the key observation is the fact that the SR gain of BD-RIS over conventional RIS significantly improves as $K$ increases due to the more flexible design of the former, at the expense of increased circuit complexity. 

Finally, the effect of varying $B$ is discussed from the results plotted in Fig. \ref{fig:bdris_b}. Although the number of DoFs are maintained in all cases, it is observed that increasing the number of scattering elements $B$ in the RIS actually increases the SR gap between RIS and BD-RIS. This is due to the fact that the RIS is placed between the transmitter and the communication users. Hence, increasing $B$ has a beneficial effect in both links. Specifically, it aids in turning the link from transmitter to RIS into a more deterministic one, and also allows for a superior spatial interference management technique in the link from RIS to users due to employing narrower and more directional beamforming as $B$ increases. Thus, this derives into the SR gap increasing along with $B$, even though the number of DoFs is maintained.  Thus, it is demonstrated that, in the large scale regime, BD-RIS offers far more superior SR performance than conventional RIS and that $B$ is the key factor in improving the overall system performance.

\begin{figure}[t!]
		\centering		        
        \includegraphics[width=\columnwidth]{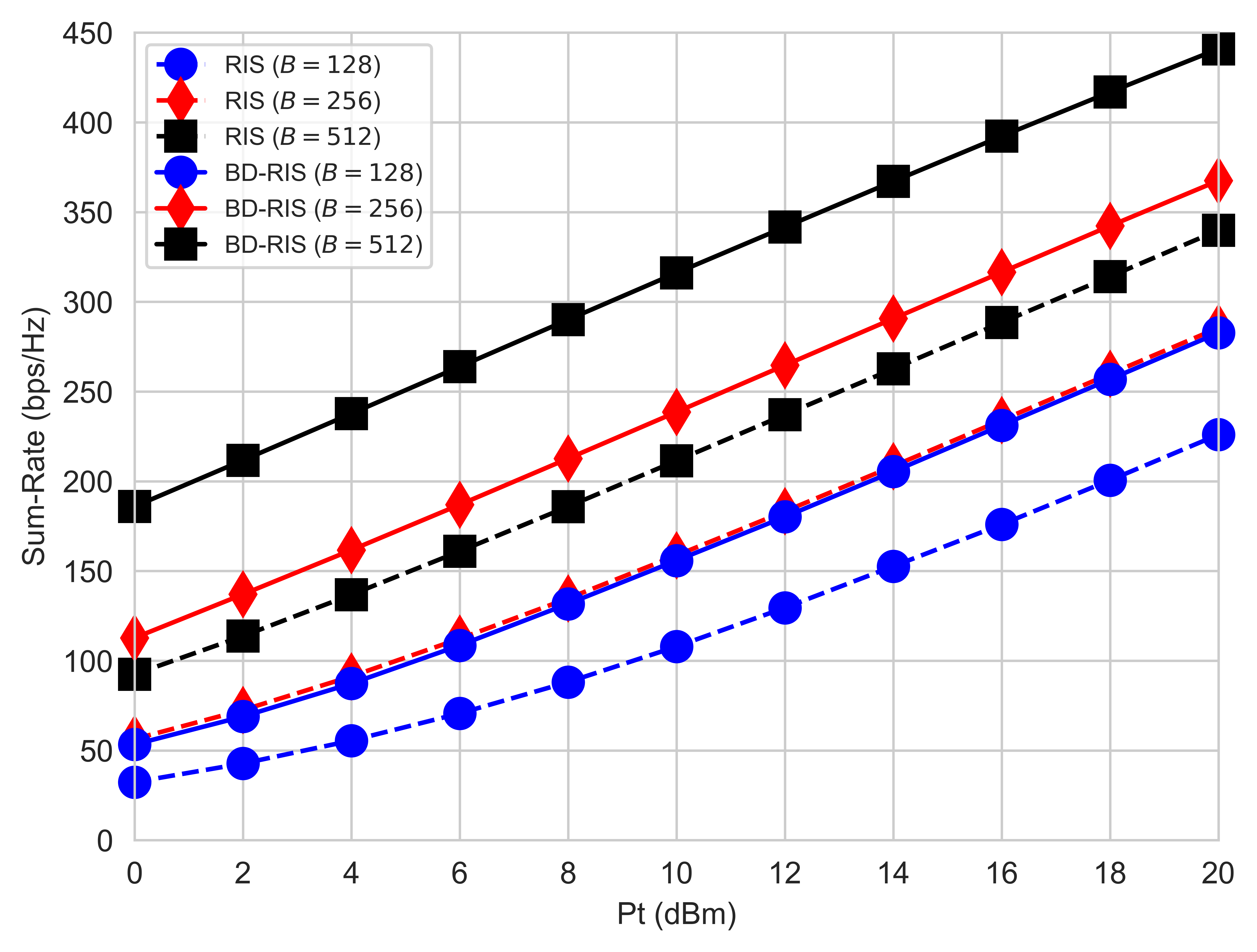}    
		\caption{ESR vs. SNR ($N_t=100, K=40$): varying $B$.}
		\label{fig:bdris_b}
\end{figure}
\section{Conclusion}
In this work, a generalized meta-learning based optimization framework is proposed to directly solve non-convex optimization problems of large scale wireless systems. Specifically, for an optimization problem with $N$ independent optimization variables, this is achieved by exploiting the overfitting effect of $N$ dedicated DNNs so that each learns to output the optimum gradient update step based on a fixed sub-optimal variable initialization. The meta-learning based optimization framework is then used to solve the ESR/SR maximization problems of three relevant, highly complex 6G settings: H-RSMA, ISAC, and BD-RIS. Numerical results reveal that the proposed meta-learning based optimization framework successfully solves the non-convex optimization problems of these three applications, and also reveals unknown aspects of their optimum operation in the large scale regime.

Future work includes the further exploration of RSMA schemes based on generalized rate-splitting \cite{eurasip}, and also the application of meta-learning algorithms to estimate the DNN structure of the proposed meta-learning based optimization framework for different deployment scales and CSIT quality levels.

\vfill\pagebreak

\end{document}